\documentclass[12pt]{article}
\usepackage{amsmath, amssymb, graphics}
\usepackage[utf8]{inputenc}
\usepackage{graphicx}
\usepackage[final]{showkeys}
\usepackage{bbm}
\usepackage{color}
\usepackage{subfigure}
\usepackage{float}
\usepackage{verbatim}
\usepackage{hep}

\begin{document}

\begin{flushright}
UUITP-05/15
\end{flushright}

\begin{center}

{\Large\bf  
Lagrangian Insertion in the Light-Like Limit and the Super-Correlators/Super-Amplitudes Duality
}

\end{center}

\vspace{\stretch{1}}

\begin{center}
 
Oluf Tang Engelund\\

\textit{Department of Physics and Astronomy, Uppsala University,}\\
\textit{SE-751 08 Uppsala, Sweden}\\
 {\tt oluf.engelund@physics.uu.se}

\end{center}

\vspace{\stretch{2}}

\begin{abstract}
In these notes we describe how to formulate the Lagrangian insertion technique in a way that mimics generalized unitarity. We introduce a notion of cuts in position space and show that the cuts of the correlators in the super-correlators/super-amplitudes duality correspond to generalized unitarity cuts of the equivalent amplitudes. The cuts consist of correlation functions of operators in the chiral part of the stress-tensor multiplet as well as other half-BPS operators. We will also discuss the application of the method to other correlators as well as non-planar contributions.
\end{abstract}
\thispagestyle{empty}
\vspace{\stretch{2}}
\newpage

\tableofcontents

\section{Introduction}

Generalized unitarity \cite{Bern:1994zx,Bern:1996je,Britto:2004nc} is a method that has been tremendously successful in computing loop-level scattering amplitudes (for a review see for instance \cite{Bern:2011qt}). It is, therefore, natural to attempt to apply a similar method to the computation of correlation functions. There are different strategies that one can employ in doing this.

One strategy is to apply generalized unitarity directly. This involves computing form factors and sewing them together to generalized unitarity cuts in momentum space. From the generalized unitarity cuts one can then construct the correlation functions in momentum space. Finally the result is Fourier transformed back into position space \cite{Engelund:2012re}. This approach has many merits: form factors of some operators have been shown to have simple structures reminiscent of the ones found in scattering amplitudes \cite{Brandhuber:2010ad,Brandhuber:2011tv}, and, although the work cited here deals with $\mathcal{N}=4$ super-Yang-Mills, it could easily be applied to other theories. Unfortunately, correlation functions are best expressed in position space so some of the symmetries may not be apparent until after the Fourier transform. Nonetheless this approach is useful, and it will be helpful to us when dealing with supersymmetry.

Our focus will be on a different strategy. We will start with a well-known position space method and try to reformulate it in a way that mimics generalized unitarity. The approach will thus not be an actual version of generalized unitarity. Rather it will be a position space method inspired by generalized unitarity. The well-known position space method useful for this strategy is the Lagrangian insertion procedure\cite{Intriligator:1998ig}. This method will be reformulated to make it similar to generalized unitarity, and a notion of cuts in position space will be introduced\footnote{In \cite{Laenen:2014jga} a slightly different notion of cuts in position space was introduced which correspond more to Cutkosky cut rules than to generalized unitarity cuts}. The advantage of this approach is that we stay in position space the whole time.

This method will be applied to the super-correlators/super-amplitudes duality. The duality relates correlation functions of operators in the chiral part of the stress-tensor multiplet to scattering amplitudes at the level of the integrands in planar $\mathcal{N}=4$ super-Yang-Mills \cite{Eden:2010zz,Eden:2010ce,Eden:2011yp,Eden:2011ku}. It was inspired by the duality between amplitudes and Wilson loops \cite{Alday:2007hr,Alday:2007he,Drummond:2007aua,Brandhuber:2007yx} whose supersymmetric version was found in \cite{Mason:2010yk,CaronHuot:2010ek}. The duality between scattering amplitudes and Wilson loops can be complicated at the quantum level because of the appearance of divergences needing to be regularized\footnote{See \cite{Belitsky:2011zm} for a discussion of some of the anomalies that this can cause}. In an attempt at clarifying matters, it was made part of a triality with correlation functions in a special light-like limit being dual to Wilson loops \cite{Alday:2010zy} and at the integrand level to scattering amplitudes \cite{Eden:2010zz,Eden:2011yp,Eden:2011ku}. In \cite{Adamo:2011dq} twistor space methods were used to prove the equivalence between the supersymmetric correlation functions and the Wilson loop introduced in \cite{Mason:2010yk}.

The super-correlators/super-amplitudes duality provides a simple example to try out our approach as one can define generalized unitarity cuts for the dual scattering amplitudes. The cuts of the correlation functions will turn out to be equivalent to the generalized unitarity cuts of the dual scattering amplitudes as long as the duality is correct in the Born approximation. The cuts will consist entirely of correlation functions of half-BPS operators whose form factors we are going to need. The calculations will not depend on the number of operators/external states in the correlation functions/amplitudes.

The duality between correlation functions and Wilson loops has also been expanded to include additional operators \cite{Alday:2011ga}. This duality has been discussed using Feynman diagram techniques in \cite{Engelund:2011fg} and using twistor space methods in \cite{Adamo:2011cd}. Even though there is no duality with scattering amplitudes, it might still be possible to compute the correlation functions with the cuts introduced here as we will discuss in the last part of the notes.

The notes are structured as follows. Section \ref{GU afsnit} deals with generalized unitarity, lists the form factors we are going to need and gives a simple example on how to use generalized unitarity for correlation functions. Section \ref{LI afsnit} deals with the Lagrangian insertion procedure and introduces the notion of position space cuts. Section \ref{Duality afsnit} deals with the duality and how to compute cuts for the correlation functions. Section \ref{General afsnit} discusses more general correlation functions and section \ref{Diskussionsafsnit} sums up the results. Note that both position space and momentum spinors appear throughout this paper: section \ref{GU afsnit} uses momentum spinors, section \ref{Duality afsnit} uses position space spinors and section \ref{susy afsnit} uses both types of spinors. This paper only considers correlation functions in $\mathcal{N}=4$ super Yang-Mills. Apart from some comments in section \ref{General afsnit}, the paper will focus exclusively on the planar theory. Though the subject of the paper is cuts in position space, we also use standard generalized unitarity cuts in momentum space. In order to distinguish properly between the two, we will always use the term 'generalized unitarity cuts' when refering to the momentum space quantities while 'cuts' will always refer to the position space quantities.

\section{Generalized Unitarity}\label{GU afsnit}

Generalized unitarity is a method for computing perturbative quantities and has been used with great success to calculate scattering amplitudes. The method exploits information found at lower loop orders by setting internal propagators on-shell. Formally, this can thought of as replacing specific propagators with delta functions:

\begin{align}
\frac{1}{p^2-m^2}&\longrightarrow \delta^{(+)}(p^2-m^2).
\end{align}

These internal propagators will then act like external states. By replacing propagators inthis way, one can eventually reduce the scattering amplitude to a product of lower order amplitudes. The product of lower order amplitudes is called a generalized unitarity cut. From the generalized unitarity cut one can reconstruct the part of the amplitude that contains the specific propagators that were replaced by delta functions. In order to compute the full amplitude, it is necessary to consider other generalized unitarity cuts until one has fully constrained the amplitude.

Since generalized unitarity explicitly refer to propagators, it depends deeply on the existence of a Feynman diagram representation. However it avoids using Feynman rules directly. Instead, on-shell amplitudes become the building blocks for the generalized unitarity cuts. This is advantageous as the on-shell amplitudes are often a lot simpler than the off-shell Feynman rules would suggest.

Generalized unitarity can also be applied to objects containing local gauge-invariant operators such as correlation functions \cite{Engelund:2012re} and form factors \cite{Brandhuber:2010ad,Brandhuber:2011tv,Bork:2011cj,Gehrmann:2011xn,Brandhuber:2012vm,Johansson:2012zv,Young:2013hda,Penante:2014sza,Nandan:2014oga}. Since generalized unitarity is a momentum space method, the local operators will have to be Fourier transformed. This introduces some off-shell momenta flowing into the generalized unitarity cuts.

In order to apply generalized unitarity to correlation functions requires form factors. Form factors are quantities in between correlation functions and amplitudes as they contain both local operators and on-shell external states. They appear because the correlation functions contain gauge-invariant operators while the method itself introduces on-shell states.

For the duality between correlation functions and scattering amplitudes, the following operators are relevant:

\begin{align}
\mathcal{T}_d(x_i,\theta^+_i)=&{}e^{\theta^{+a}_{i\alpha} Q_{i+a}^\alpha}\mathrm{Tr}\left((\phi^{++})^d\right).\label{def T_k}
\end{align}

Here harmonic variables have been used to make the following projections:

\begin{align}
\theta_{i\alpha}^{\pm a}=&{}\theta^A_{i\alpha}(i)^{\pm a}_A,&Q_{i\pm a}^\alpha=&{}Q_A^\alpha(\bar{\imath})^A_{\pm a},&\phi^{++}=&{}-\tfrac{1}{2}\phi^{AB}(i)^{+a}_A\epsilon_{ab}(i)^{+b}_B,
\end{align}
\noindent of the super space, super charges and scalar fields respectively. In the above $a,b$ are SU(2) indices, $\alpha$ is a spinor index and $A,B$ are the usual R-symmetry indices. We will follow the notation and conventions of \cite{Eden:2011yp,Eden:2011ku} closely with respect to both harmonic variables and spinors. Some of the conventions can be found in appendix \ref{notationsappendix}.

The form factors for these operators are very simple as the operators respect part of the supersymmetry. They have been dealt with extensively in the papers \cite{Brandhuber:2011tv,Penante:2014sza}. For our purposes we are only going to need MHV form factors as we will explain later. For $d=2$ the super-Fourier transform of the MHV form factor is given by:

\begin{align}
\mathcal{F}^{\mathrm{MHV}}_{\mathcal{T}_2}(\gamma_{i+a}^\alpha,1,\cdots,n)=&{}\frac{\delta^8\left((i)^{+a}_{A}\gamma_{i+a}^\alpha-\sum_{r=1}^n\eta_{rA}\lambda_r^\alpha\right)}{\langle12\rangle\langle23\rangle\cdots \langle n1\rangle}.\label{F_T_2}
\end{align}

This particular operator is part of the stress-tensor multiplet. Its highest component is the on-shell chiral Lagrangian that will also appear as part of the Lagrangian insertion procedure:

\begin{align}
\mathcal{T}_2(x_i,\theta_i^+)=&{}\mathrm{Tr}\left(\phi^{++}\phi^{++}\right)+\cdots+\frac{1}{3}(\theta_i^+)^4\mathcal{L}(x_i)\label{T_2}
\end{align}

In order to write \eqref{F_T_2} in terms of the super-space variables one has to do an inverse super-Fourier transform:

\begin{align}
\mathcal{F}(\theta_{i\alpha}^{+a},x_i,1,\cdots,n)=&{}\int \frac{d^4q}{(2\pi)^4}d^4\gamma e^{iq_i\cdot x_i+i\theta_{i\alpha}^{+a}\gamma_{i+a}^\alpha}\delta^4\left(q_i-\sum_{r=1}^n p_r\right)\mathcal{F}(\gamma_{i+a}^\alpha,1,\cdots,n),
\end{align}
\noindent so the on-shell chiral Lagrangian correspond to the part of \eqref{F_T_2} proportional to $(\gamma)^0$.

For $d>2$ MHV form factors will have a fermionic content that, in addition to the super-momentum conserving delta function. If we define the quantity $\widetilde{\mathcal{F}}_{\mathcal{T}_d}$ as the form factor excluding the super-momentum conserving delta function:

\begin{align}
\mathcal{F}_{\mathcal{T}_d}(\gamma_{i+a}^\alpha,1,\cdots,n)=&{}\widetilde{\mathcal{F}}_{\mathcal{T}_d}(1,\cdots,n)\delta^8\left((i)^{+a}_{A}\gamma_{i+a}^\alpha-\sum_{r=1}^n\eta_{rA}\lambda_r^\alpha\right),
\end{align}

\noindent then $\widetilde{\mathcal{F}}^{\rm MHV}_{\mathcal{T}_d}$ will be a polynomial of degree $2(d-2)$ in $\eta_{-a}=(\bar{\imath})_{-a}^A\eta_A$. Some interesting relations between the form factors for an operator $\mathcal{T}_d$ and form factors for an operator $\mathcal{T}_{d-1}$ were found in \cite{Penante:2014sza} using BCFW recursion. However we are not interested in the explicit expressions for $\widetilde{\mathcal{F}}$. We only need to know its degree, and that it conatins non-zero terms with $d-2$ factors of $\eta_{i-a}\epsilon^{ab}\eta_{i-b}$ for any set of $i$'s. The second fact follow from simple Feynman diagrams as there is always a non-zero form factor for $\mathrm{Tr}((\phi^{++})^d)$ with $d$ external scalars and any number of positive helicity gluons regardless of the ordering of the external states. Conservation of super momentum can then be used to make $\widetilde{\mathcal{F}}$ independent of two of the $\eta_-$'s.

The $\overline{\rm MHV}$ form factor can be written as follows:

\begin{align}
\mathcal{F}^{\overline{\mathrm{MHV}}}_{\mathcal{T}_2}(\gamma_{i+a}^\alpha,1,\cdots,n)=&{}\frac{\delta^4\left(\gamma_{i+a}^\alpha-(\bar{\imath})^A_{-a'}\sum_{r=1}^n\eta_{rA}\lambda_r^\alpha\right)}{[12][23]\cdots [n1]}\\
&\int \left(\prod_{j=1}^nd^4\tilde{\eta}_j\right)e^{i\sum_{j=1}^n\eta_{jA}\tilde{\eta}^A_j}\delta^4\left((\bar{\imath})^A_{+a}\sum_{r=1}^n\eta_{rA}\lambda_r^\alpha\right).\nonumber
\end{align}

After performing the $\tilde{\eta}$-integrations, this formula becomes a Grassmann polynomial of degree $4n$. For the case $n=2$, it is equivalent to the MHV formula while for $n>2$ it is a Grassmann polynomial of a higher degree than the MHV formula. Unlike for scattering amplitudes where the three-point $\overline{\rm MHV}$ amplitude is a Grassmann polynomial of only degree 4, there are no special form factors for the operators in \eqref{T_2} with a lower degree than the MHV formula.

\begin{figure}
\begin{center}
\begin{tabular}{cc}
\includegraphics[scale=0.8]{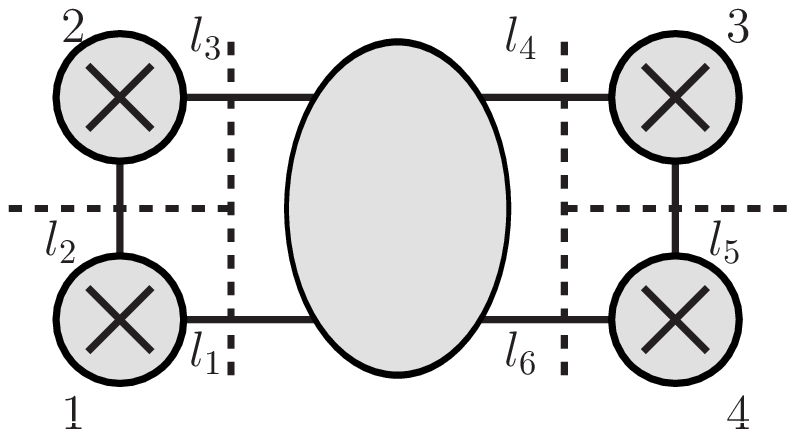}&\includegraphics[scale=0.8]{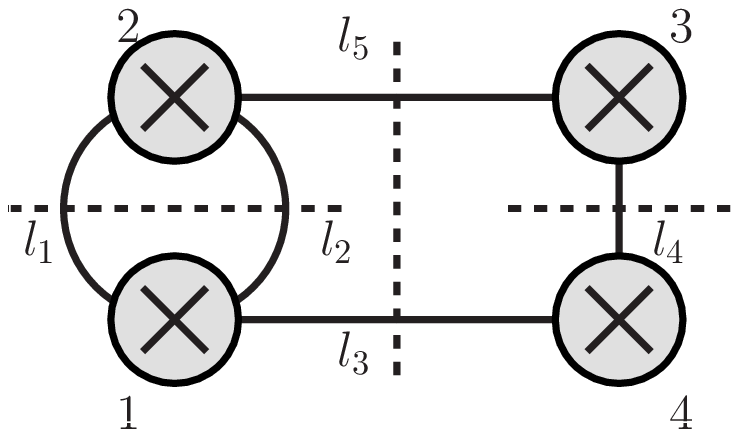}\\
(a)&(b)\\
\includegraphics[scale=0.8]{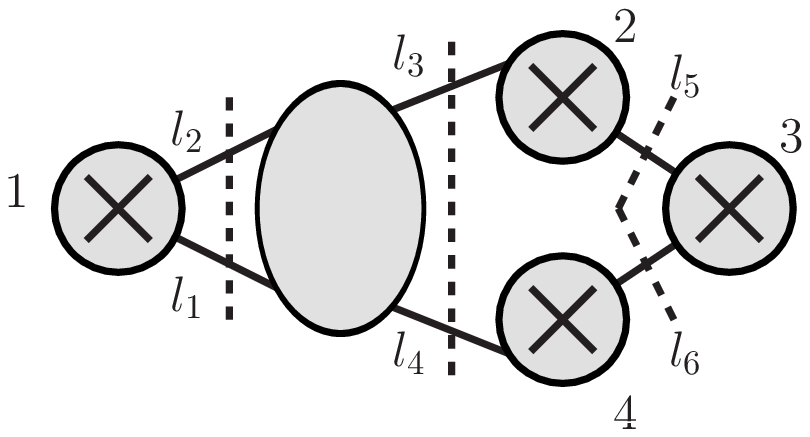}&\includegraphics[scale=0.8]{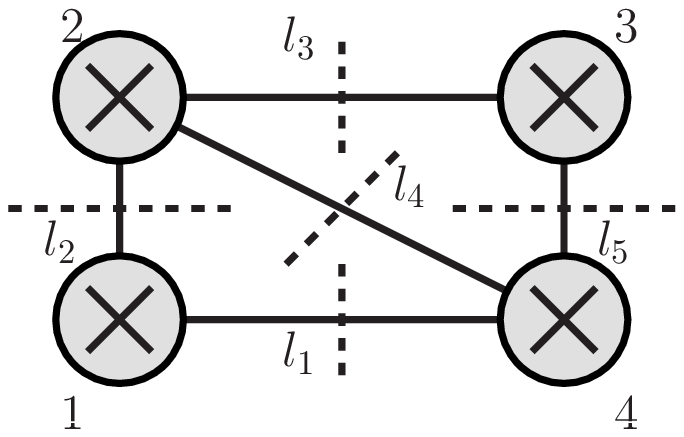}\\
(c)&(d)
\end{tabular}
\end{center}
\caption{Four generalized unitarity cuts where the crosses indicate the quantities are form factors while the blobs without crosses are scattering amplitudes. The numbers $i$ by the form factors indicate that the operator is placed at the point $x_i$. In cut (b) the form factor associated with the operator at $x_2$ is $\overline{\rm MHV}$ while in cut (d) the form factor associated with the operator at $x_4$ is $\overline{\rm MHV}$. The remaining quantities are MHV\label{Eksempel cuts}}
\end{figure}

As an example of how to use generalized unitarity on correlation functions, consider the correlator of four operators $\mathrm{Tr}(\phi^{++}\phi^{++})$ placed at four different locations. This correlation function can be computed using the four cuts shown in figure \ref{Eksempel cuts} as well as those with the locations of the operators permuted. In the diagrams, the blobs with crosses are form factors while the blobs without are amplitudes.

The generalized unitarity cuts will written in terms of spinors and products of harmonic variables defined as follows:

\begin{align}
\mathbf{(ij)}&=\tfrac{1}{4}\epsilon^{ABCD}(i)^{+a}_{A}\epsilon_{ab}(i)^{+b}_{B}(j)^{+c}_{C}\epsilon_{cd}(j)_{D}^{+d}.
\end{align}

The generalized unitarity cuts can be found to be:

\begin{align}
\mathrm{Cut}_a=&{}{-}2N_c(N_c^2-1){\bf (12)(34)}\left(-{\bf (12)(34)}+{\bf (23)(14)}\frac{\langle l_1l_3\rangle\langle l_4l_6\rangle}{\langle l_4l_3\rangle\langle l_6l_1\rangle}+{\bf (13)(24)}\frac{\langle l_1l_3\rangle\langle l_6l_4\rangle}{\langle l_1l_4\rangle\langle l_3l_6\rangle}\right)\\
\mathrm{Cut}_b=&{}{-}2N_c(N_c^2-1)\frac{\bf (12)(23)(34)(14)}{(l_1+l_2)^2}\left(\frac{[l_3l_1]\langle l_1l_5\rangle}{[l_3l_2]\langle l_2 l_5\rangle}+\frac{[l_3l_2]\langle l_2l_5\rangle}{[l_3l_1]\langle l_1l_5\rangle}+2\right)\\
\mathrm{Cut}_c=&{}{-}2N_c(N_c^2-1){\bf (12)(23)(34)(14)}\langle l_3l_4\rangle\langle l_1l_2\rangle\left(\frac{1}{\langle l_3l_2\rangle\langle l_4l_1\rangle}-\frac{1}{\langle l_1l_3\rangle\langle l_2l_4\rangle}\right)\\
\mathrm{Cut}_d=&{}{-}2N_c(N_c^2-1){\bf (12)(23)(34)(14)}\frac{[l_1l_5]\langle l_3l_2\rangle}{[l_1l_4][l_4l_5]\langle l_3l_4\rangle\langle l_4l_2\rangle}
\end{align}

To make the result as transparent as possible we define functions $a$ and $b$ such that the one-loop result can be written as:

\begin{align}
\langle T_2(x_1,0)&T_2(x_2,0)T_2(x_3,0)T_2(x_4,0)\rangle^{(1)}\nonumber\\
=&{}-2(N_c^2-1)\left(\frac{g^2N_c}{4\pi^2}\right)\bigg[{\bf (12)}^2{\bf (34)}^2a(1,2)+{\bf (13)}^2{\bf (24)}^2a(1,3)+{\bf (14)}^2{\bf (23)}^2a(1,4)\label{4-pt opdeling}\\
&+{\bf (12)(23)(34)(14)}b(1,2,3,4)+{\bf (12)(24)(34)(13)}b(1,2,4,3)\nonumber\\
&+{\bf (13)(23)(24)(14)}b(1,3,2,4)\bigg]\nonumber.
\end{align}

\begin{figure}
\begin{center}
\begin{tabular}{cccc}
\includegraphics[scale=0.5]{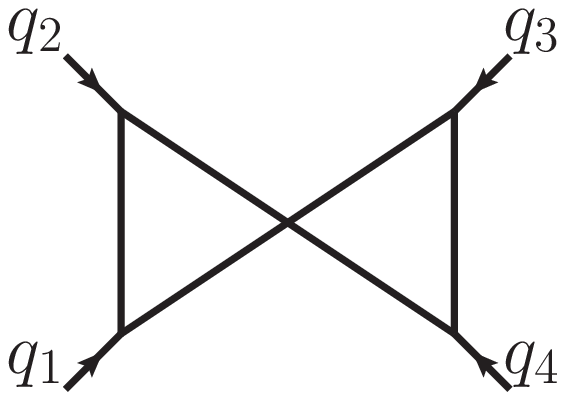}&\includegraphics[scale=0.5]{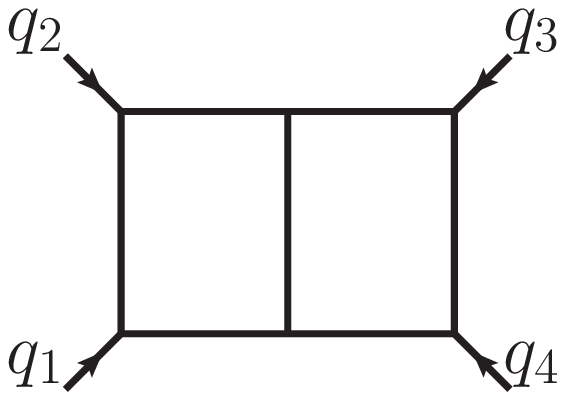}&\includegraphics[scale=0.5]{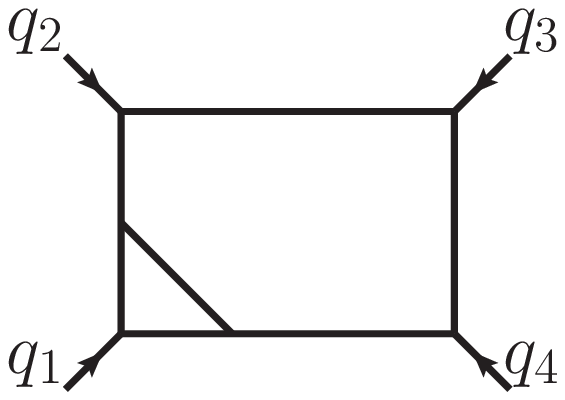}&\includegraphics[scale=0.5]{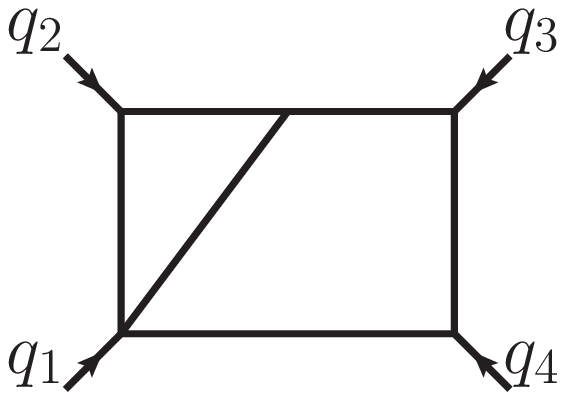}\\
${\rm BTie}(1,2|3,4)$&${\rm DB}(1,2|3,4)$&${\rm TriP}(1|2,3,4)$&${\rm TriB}(1|2|3,4)$
\end{tabular}
\end{center}
\caption{Integrals used for the four-point example. The $q_i$'s are off-shell momenta associated with the gauge-invariant operators at the points $x_i$}
\label{integraler}
\end{figure}

The Fourier transforms of these functions can then be determined from the above generalized unitarity cuts. Written in terms of the scalar integrals from figure \ref{integraler}, they are given by:

\begin{align}
\tilde{a}(1,2)=&{}-{\rm BTie}(1,2|3,4),\\
\tilde{b}(1,2,3,4)=&{}(q_1+q_2)^2{\rm DB}(1,2|3,4)+(q_1+q_4)^2{\rm DB}(4,1|2,3)+q_1^2{\rm TriP}(1|2,3,4)\label{b-tilde}\\
&+q_2^2{\rm TriP}(2|3,4,1)+q_3^2{\rm TriP}(3|4,1,2)+q_4^2{\rm TriP}(4|1,2,3)-{\rm TriB}(1|2|3,4)\nonumber\\
&-{\rm TriB}(2|3|4,1)-{\rm TriB}(3|4|1,2)-{\rm TriB}(4|1|2,3)-{\rm TriB}(4|3|2,1)\nonumber\\
&-{\rm TriB}(3|2|1,4)-{\rm TriB}(2|1|4,3)-{\rm TriB}(1|4|3,2)\nonumber.
\end{align}

These results can be written in position space as an integral over a single space-time point $y$. This transformation is relatively simple for the ${\rm BTie}$ and the ${\rm TriB}$ integrals as they only have a single interaction vertex apart from those related to the gauge invariant operators. By writing momentum conservation at this vertex as the integration over a space-time point, the integrals simply become a collection of propagators connecting the different points. The function $a$ can this way be written as:

\begin{align}
a(1,2)=&{}\frac{1}{(4\pi^2)^5}\frac{1}{(x_1-x_2)^2(x_3-x_4)^2}\int \frac{d^4y}{(x_1-y)^2(x_2-y)^2(x_3-y)^2(x_4-y)^2}.
\end{align}

The other integrals are a bit more complicated to Fourier transform. However it is possible to rewrite the expression using relations for the Fourier transforms of these integrals. The relations can be found in \cite{Engelund:2012re,Eden:1998hh,Beisert:2002bb}\footnote{Equation (C.17) in \cite{Engelund:2012re}}, and with those it is possible to write the $b$ function as:

\begin{align}
b(1,2)=&{}\frac{1}{(4\pi^2)^5}\frac{(x_1-x_3)^2(x_2-x_4)^2-(x_1-x_2)^2(x_3-x_4)^2-(x_1-x_4)^2(x_2-x_3)^2}{(x_1-x_2)^2(x_2-x_3)^2(x_3-x_4)^2(x_1-x_4)^2}\label{b}\\
&\int \frac{d^4y}{(x_1-y)^2(x_2-y)^2(x_3-y)^2(x_4-y)^2}\nonumber.
\end{align}

Generalized unitarity does not seem to be as effective when applied to correlation functions as to scattering amplitudes. The issue is that correlation functions are best formulated in position space, whereas generalized unitarity is a method that must be applied in momentum space. Indeed the simplicity of \eqref{b} is in no way apparent in the momentum space expression from equation \eqref{b-tilde}. Nonetheless, this technique can be very useful, and we will employ it when dealing with the wholly supersymmetric case. Although in this case, we will first use the Lagrangian insertion procedure so generalized unitarity is applied to a Born-level correlator, more on this in section \ref{GU on LI}.

\section{Lagrangian Insertion in the Light-Like Limit}\label{LI afsnit}

Lagrangian insertion is a useful method for constructing correlation functions in $\mathcal{N}=4$ super-Yang-Mills. It exploits the fact that, after a suitable rescaling of fields, differentiation of a correlation function with respect to the coupling will bring down a factor of the on-shell chiral Lagrangian:

\begin{align}
\mathcal{L}(x)=&{}\mathrm{Tr}\left(-\tfrac{1}{2}F_{\alpha\beta}F^{\alpha\beta}+\sqrt{2}g\psi^{\alpha A}[\phi_{AB},\psi^B_\alpha]-\tfrac{1}{8}g^2[\phi^{AB},\phi^{CD}][\phi_{AB},\phi_{CD}]\right).\label{on-shell Lagrangian}
\end{align}

This operator also appeared in the expansion of the operator $\mathcal{T}_2$ in \eqref{T_2}. The trick allows one to relate the $l$th order correction of the correlator:

\begin{align}
\langle\mathcal{O}(x_1)\cdots\mathcal{O}(x_n)\rangle,\label{correlator without Ls}
\end{align}
\noindent to the $l-m$th order correction of the correlator:

\begin{align}
\int d^4y_1\cdots d^4y_m\langle\mathcal{O}(x_1)\cdots\mathcal{O}(x_n)\mathcal{L}(y_1)\cdots\mathcal{L}(y_m)\rangle.\label{correlator with Ls}
\end{align}

When computing the correlator in \eqref{correlator with Ls}, we can neglect contact terms $i.e.$ terms proportional to a space-time delta function. In general, we ignore terms proportional to delta functions of the type:

\begin{align*}
&\delta^4(x_i-x_j),
\end{align*}
\noindent as the original operators are all placed at different locations in the correlation functions relevant to the duality. Part of the Lagrangian insertion procedure is to also ignore terms including delta functions of the types:

\begin{align*}
&\delta^4(x_i-y_j),&&\delta^4(y_i-y_j).
\end{align*}

Terms with such delta functions will in \eqref{correlator with Ls} act like terms from the lower loop orders. Because of the rescaling of fields, the derivative with respect to the coupling constant could also act on the operators themselves. These terms will similarly act as if they were of a lower loop order. It has been argued that these two types of terms cancel out (see for instance \cite{CaronHuot:2010ek}). There does not seem to be a formal proof for this in general, but we will assume that it holds, and it will be important to some of the later arguments.


In addition to being easier than a direct application of Feynman rules, Lagrangian insertion also gives the correlator in a form that mimics more closely the form that scattering amplitudes have in momentum space. Notice for instance that after using the Lagrangian insertion procedure to relate the original correlator to a Born-level correlation function, the $l$th order correction will naturally contain $l$ variables to be integrated over. This is similar to the way that the loop order $l$ of scattering amplitudes contain $l$ loop momenta.

Normally one would compute the correlator in \eqref{correlator with Ls} using standard Feynman rules but inspired by generalized unitarity, we will instead consider different limits of the type:

\begin{align}
\lim_{\begin{array}{c}
(y_i-y_j)^2=0\\
(x_i-y_j)^2=0
\end{array}}\frac{\langle\mathcal{O}(x_1)\cdots\mathcal{O}(x_n)\mathcal{L}(y_1)\cdots\mathcal{L}(y_m)\rangle}{\langle\mathcal{O}(x_1)\cdots\mathcal{O}(x_n)\mathcal{O}(y_1)\cdots\mathcal{O}(y_m)\rangle^{(0)}},\label{cut def}
\end{align}
\noindent where each limit consists of a set of distances becoming light-like. In the denominator the Lagrangian insertions have to be replaced by other operators since the Lagrangians cannot be connected directly to each other but only by going through vertices so the lowest non-zero correlator would be at some loop level. The relevant operators will be the lowest fermionic components of the operators described in section \ref{GU afsnit} as we want the denominator to just be a collection of scalar propagators. The light-like distances fall into three different categories: $y_i-y_j$, $y_i-x_j$ and $x_i-x_j$ though we will mainly be interested in the first two types. The last type would be important for a BCFW recursion relation \cite{Britto:2004ap,Britto:2005fq}\footnote{This last type of limit has been used in \cite{Eden:2011yp,CaronHuot:2010ek,Adamo:2011cd}}.

Similarly to generalized unitarity no limit will give the full result but each limit will determine a specific part of the full expression, and one will have to compute several different limits until the integrand is completely fixed. It is of course not immediately obvious that these limits will completely determine the integrand, or to borrow an expression from generalized unitarity that the correlation function is cut-constructible. The correlation functions relevant to the super-correlators/super-amplitudes duality are however cut-constructible. This follows from the operator product expansion.

The existence of an operator product expansion ensure that the correlator should be a function of differences between space-time points (say $(x_i-x_j)^2$). For the operators in the chiral part of the stress-tensor multiplet \eqref{T_2}, the operator product expansion also imply that the correlation functions at the Born level will contain only poles of order one and two. The poles of order two come from disconnected graphs which we are not interested in. Above the Born level, the correlation function could also contain logarithms of the differences of two points \cite{Eden:2011yp,Eden:2011we}. The presence of logarithms would make it harder to argue in favor of cut-constructibility. Therefore, we will always use the Lagrangian insertion procedure in such a way that the correlator in equation \eqref{correlator with Ls} is at the Born level. As the duality only deals with connected graphs, this is enough to ensure that the correlation functions are cut-construtible.


\section{The Super-Correlators/Super-Amplitudes Duality}\label{Duality afsnit}

The duality between correlation functions and scattering amplitudes considers operators of the type \eqref{T_2} placed at points $x_1$ to $x_n$ with neighbouring points being light-like separated but otherwise generic, thereby creating a polygon. The sides of the polygon are identified with on-shell momenta:

\begin{align}
p_i^{\alpha\dot{\alpha}}=&{}(x_i-x_{i+1})^{\alpha\dot{\alpha}}=\lambda_i^\alpha\tilde{\lambda}_i^{\dot{\alpha}},\label{real space spinors}
\end{align}
\noindent while the superspace variables are identified with the fermionic parts of the supertwistor:

\begin{align}
\chi_{i/i}^a\equiv\langle i\theta_i^{+a}\rangle=&{}\chi_i^A(i)^{+a}_A&\chi_{i/i+1}^a\equiv\langle i\theta_{i+1}^{+a}\rangle=&{}\chi_i^A(i+1)^{+a}_A\label{fermionic identification}
\end{align}

The duality considers the ratio of the connected part of the correlation function over its Born-level expression. This is then equal to the square of a color-ordered amplitude divided by its tree-level MHV formula:

\begin{align}
\lim_{(x_i-x_{i+1})^2=0}\frac{G_n}{G_n^{(0)}}=&{}\left(\sum_k\left(\frac{g^2N_c}{4\pi^2}\right)^k\frac{A^{\mathrm{N^kMHV}}_n}{A_n^{\mathrm{MHV}(0)}}\right)^2,\label{duality eq}
\end{align}
\noindent where $g$ is the coupling constant. An $\mathrm{N^kMHV}$ amplitude will correspond to $4k$ factors of the super-space variables on the correlator side of the duality but the lowest non-trivial order of a correlation function with that many super-space variables is proportional to $g^{2k}$. This is the reason behind the factor dependent on the coupling constant.

Our goal is to describe how to compute the correlator on the left-hand side of the equation through position space cuts. Those cuts will turn out to be equivalent to the generalized unitarity cuts for the amplitude on the right-hand side.

\subsection{Divergences and Wilson Lines}\label{div Wilson}

\begin{figure}
\begin{center}
\includegraphics[scale=0.8]{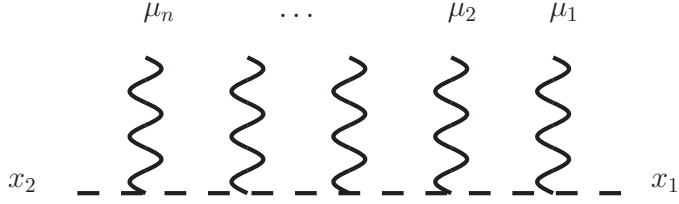}
\end{center}
\caption{Two scalars connected through a sequence of single-gluon vertices\label{WLg}}
\end{figure}

Before we proceed to consider the light-like limits involving Lagrangian insertions, let us briefly summarize some conclusions from \cite{Engelund:2011fg}. They will be important in the later sections. In that paper, the light-like limit of single distances were analyzed using Feynman rules. One of the examples was two scalar fields connected through a sequence of $n$ single-gluon vertices, as shown in figure \ref{WLg}. If we denote the momenta flowing out from the endpoints by $q_1$ and $q_2$, the momenta flowing in with the gluons by $p_j$ and the momenta of the scalar propagators by $k_j$, the expression can be written as follows:

\begin{align}
L_{\mu_1\cdots\mu_n}(x_1,x_2,p_1\cdots p_n)=&{}\int \frac{d^D q_1}{(2\pi)^D}\frac{d^Dq_2}{(2\pi)^D}e^{iq_1\cdot x_1+iq_2\cdot x_2}\int \frac{d^D k_1}{(2\pi)^D}\cdots \frac{d^D k_{n-1}}{(2\pi)^D}\label{gammelt1}\\
&\times\frac{(q_1-k_1)_{\mu_1}(-k_1-k_2)_{\mu_2}\cdots (-k_{n-1}-p_2)_{\mu_n}}{(q_1^2+i\epsilon)\prod_{j=1}^{n-1}(k_j^2+i\epsilon)(q_2^2+i\epsilon)}(2\pi)^D\delta^D(p_1+k_1+q_1)\nonumber\\
&\times(2\pi)^D\delta^D(p_2-k_1+k_2)\cdots(2\pi)^D\delta^D(p_{n-1}-k_{n-2}+k_{n-1})\nonumber\\
&\times(2\pi)^D\delta^D(p_n-k_{n-1}+q_2)\nonumber.
\end{align}
\noindent where $\epsilon>0$. By introducing Feynman parameters and performing the momentum integrals this expression can be rewritten as:

\begin{align}
L_{\mu_1\cdots\mu_n}(x_1,x_2,p_1\cdots p_n)=&{}\frac{(-i)^{n+1}}{(2\pi)^D}\prod_{j=1}^n\left(-2i\frac{\partial}{\partial x_1^{\mu_j}}+2\sum_{r=1}^{j-1}p_{r\mu_j}+p_{j\mu_j}\right)\nonumber\\
&\times\left(\prod_{j=1}^n\int_{t_{j-1}}^1dt_j\right)e^{-i\sum_{j=1}^np_j\cdot (x_2t_j+x_1(1-t_j))}
\label{gammelt2}\\
&\int_0^\infty d\zeta\zeta^n\frac{\pi^{D/2}}{(-i\zeta)^{D/2}}e^{i\zeta f(t_j,p_j)-i\frac{(x_1-x_2)^2}{4\zeta}-\epsilon\zeta-\epsilon/\zeta},\nonumber
\end{align}
\noindent with $f(t_j,p_j)$ being a function whose specific expression is irrelevant. For this to be as divergent as a single scalar propagator in the light-like, the factor $\zeta^n$ in the $\zeta$-integral will have to be removed. This means that only the term involving all $n$ derivatives will survive in the light-like limit. The derivatives will become proportional to $(x_1-x_2)^\mu$, and the expression will act like a single Wilson line. In general, it was concluded that the number of derivatives minus the number of propagators decided how divergent a side was.

As the operator in equation \eqref{T_2} includes both fermions and field strenghts, one might expect the correlation functions in the duality to contain something more divergent than simple scalar propagators. However due to the chirality of the operator this is not the case. Only the scalars in the operators $T_2$ can connect through free propagators. The other fields have to connect through interaction vertices that will lower the divergences to that of simple scalar propagators. Because of this, the Born-level correlator $G_n^{(0)}$ in \eqref{duality eq} is just a collection of scalar propagators while the full correlation function $G_n$ do not become more divergent than $G_n^{(0)}$. Lagrangian insertions can be included in this analysis as the chiral on-shell Lagrangian is simply the highest component of the operator \eqref{T_2}. So we should not get anything more divergent than scalar propagators. This also matches the conclusions referenced in section \ref{LI afsnit} coming from the operation product expansion.

We will use the approach of \cite{Engelund:2011fg} in the case of a purely scalar polygon where it will provide some clear insight. We will not use it for the supersymmetric case because it becomes rather cumbersome, especially finding the correct fields that sit at the corners of the polygon. The sides of the polygon do seem to act like the supersymmetric Wilson loops of \cite{Mason:2010yk,CaronHuot:2010ek} but the appearance of ghosts at higher loop orders complicates matters.

\subsection{Combining with Generalized Unitarity}\label{GU on LI}

In the following sections, we will use regular generalized unitarity in momentum space when investigating the position space cuts. Let us consider, what kind of generalized unitarity cuts will be interesting.

Consider two operators placed at the points, $y_1$ and $y_2$. Associate the off-shell momenta $q_1$ and $q_2$ with the Fourier-transforms of these operators. If we denote the correlation function in momentum space by $C(q_1,q_2,\cdots,q_n)$ then transforming back to position space works as follows:

\begin{align}
\int d^4q_1d^4q_2e^{iq_1\cdot y_1+iq_2\cdot y_2}\delta^4\left(\sum_{j=1}^nq_j\right)&C(q_1,q_2,\cdots,q_n)\label{cut imellem op}\\
&=\int d^4q_1e^{iq_1\cdot (y_1-y_2)-i\sum_{j=3}^nq_j\cdot y_2}C(q_1,-q_1-\sum_{j=3}^nq_j,\cdots,q_n)\nonumber.
\end{align}

If $C(q_1,-q_1-\sum_{j=3}^nq_j,\cdots,q_n)$ contains no propagators between the point $y_1$ and the point $y_2$, the integral over $q_1$ will create a delta function $\delta^4(y_1-y_2)$, possibly with some additional derivatives with respect to $y_1$. We can then use the fact that any contact terms are thrown away as part of the Lagrangian insertion procedure. This means that we only need to consider generalized unitarity cuts with single-operator form factors. 

In section \ref{LI afsnit}, we described the Lagrangian insertion procedure in a generic way, where a loop level correlator was related to a correlator of a lower loop level. As mentioned in section \ref{GU afsnit}, we will always relate the loop level correlator to a Born level correlator. This has consequences for the generalized unitarity cuts, we will use.

Consider the $l$th order correction to a super-correlator for which the total number of superspace variables sum up to $4k$. This specific loop order can be computed by using $l$ Lagrangian insertions, and it should be proportional to the coupling constant to the power $2l+2k$. As argued above, it is sufficient to consider generalized unitarity cuts with only single-operator form factors. We will therefore consider the generalized unitarity cuts with just enough cut propagators to ensure, that there are no form factors with two or more gauge-invariant operators. The form factors for the operator $\mathcal{T}_2$ with $2c_i$ external legs are proportional to the coupling constant to the power $2(c_i-1)$ at tree level. This can be combine with the above way of counting the power of the coupling constant to give the relation:

\begin{align}
2\sum_i(c_i-1)=&{}2l+2k.
\end{align}

Since there are $n+l$ operators, this can be rewritten as:

\begin{align}
\sum_ic_i=&{}n+k+2l,
\end{align}
\noindent and because there are no external legs the sum over the $c_i$'s will be equal to the number of cut propagators. Every cut propagator comes with an integration over the Grassmann variables, and so the form factors should have exactly $4(n+k+2l)$ Grassmann variables. This can be accomplished if they are all MHV, and as mentioned earlier there are no $\overline{\rm MHV}$ form factors with less Grassmann variables than the MHV form factors. In total this means that in order to compute the relevant correlation functions, it is sufficient to consider generalized unitarity cuts with only MHV single-operator form factors.

\begin{figure}
\begin{center}
\begin{tabular}{cc}
\includegraphics[scale=0.8]{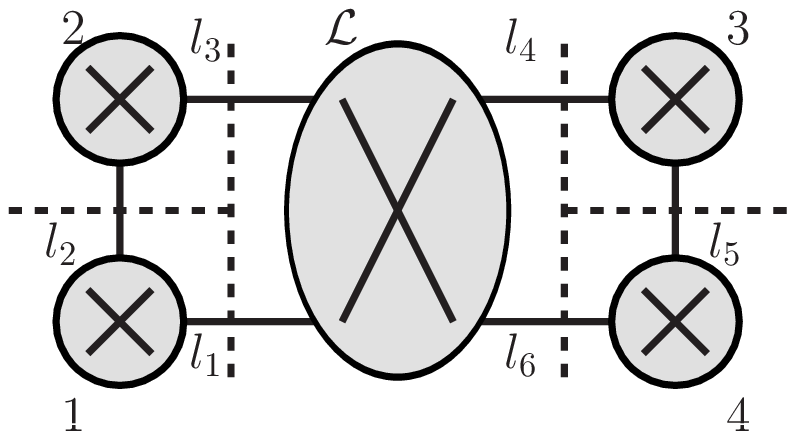}&\includegraphics[scale=0.8]{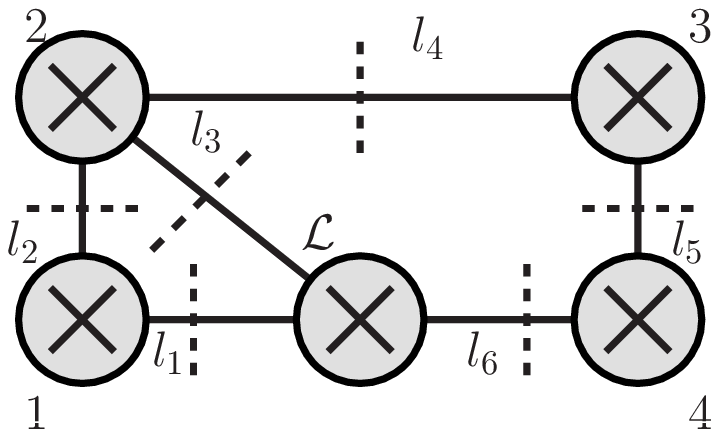}\\
(A)&(B)\\
\includegraphics[scale=0.8]{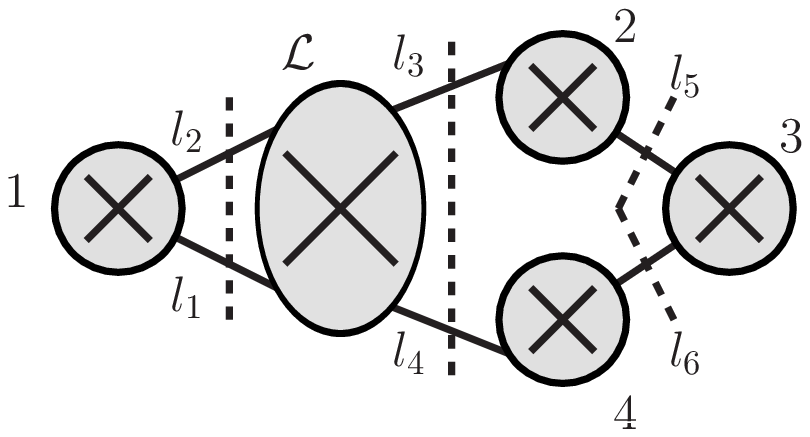}&\includegraphics[scale=0.8]{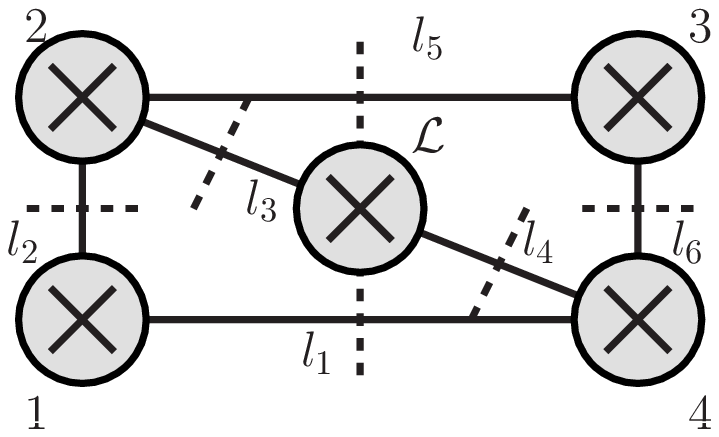}\\
(C)&(D)\\
\multicolumn{2}{c}{\includegraphics[scale=0.8]{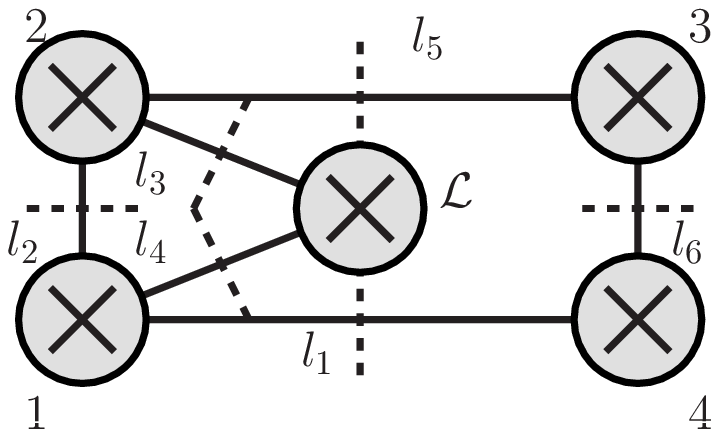}}\\
\multicolumn{2}{c}{(E)}
\end{tabular}
\end{center}
\caption{Generalized unitarity cuts used to compute the 4-point correlation function\label{AndetEksempel cuts}}
\end{figure}

As an example, consider the same correlator we computed in section \ref{GU afsnit}. This correlator can be computed from the generalized unitarity cuts shown in figure \ref{AndetEksempel cuts}. In each generalized unitarity cuts one of the form factors is of an on-shell Lagrangian (indicated in the diagrams by an $\mathcal{L}$). These generalized unitarity cuts are given by:

\begin{align}
\mathrm{Cut}_A=&{}{-}2N_c(N_c^2-1){\bf (12)(34)}\left(-{\bf (12)(34)}+{\bf (23)(14)}\frac{\langle l_1l_3\rangle\langle l_4l_6\rangle}{\langle l_4l_3\rangle\langle l_6l_1\rangle}+{\bf (13)(24)}\frac{\langle l_1l_3\rangle\langle l_6l_4\rangle}{\langle l_1l_4\rangle\langle l_3l_6\rangle}\right)\\
\mathrm{Cut}_B=&{}{-}2N_c(N_c^2-1){\bf (12)(23)(34)(14)}\left(\frac{\langle l_1l_3\rangle\langle l_4l_6\rangle}{\langle l_6l_1\rangle\langle l_3l_4\rangle}-\frac{\langle l_1l_3\rangle\langle l_2l_6\rangle}{\langle l_6l_1\rangle\langle l_3l_2\rangle}\right)\\
\mathrm{Cut}_C=&{}{-}2N_c(N_c^2-1){\bf (12)(23)(34)(14)}\langle l_3l_4\rangle\langle l_1l_2\rangle\left(\frac{1}{\langle l_3l_2\rangle\langle l_4l_1\rangle}-\frac{1}{\langle l_1l_3\rangle\langle l_2l_4\rangle}\right)\\
\mathrm{Cut}_D=&{}{-}2N_c(N_c^2-1){\bf (12)(23)(34)(14)}\left(\frac{\langle l_2l_6\rangle\langle l_3l_4\rangle}{\langle l_3l_2\rangle\langle l_4l_6\rangle}+\frac{\langle l_1l_2\rangle\langle l_3l_4\rangle}{\langle l_2l_3\rangle\langle l_1l_4\rangle}+\frac{\langle l_1l_5\rangle\langle l_3l_4\rangle}{\langle l_3l_5\rangle\langle l_1l_4\rangle}+\frac{\langle l_3l_4\rangle\langle l_5l_6\rangle}{\langle l_5l_3\rangle\langle l_4l_6\rangle}\right)\\
\mathrm{Cut}_E=&{}{-}2N_c(N_c^2-1){\bf (12)(23)(34)(14)}\left(\frac{\langle l_1l_2\rangle\langle l_3l_4\rangle}{\langle l_2l_3\rangle\langle l_4l_1\rangle}+\frac{\langle l_1l_5\rangle\langle l_3l_4\rangle}{\langle l_3l_5\rangle\langle l_4l_1\rangle}+\frac{\langle l_2l_5\rangle\langle l_3l_4\rangle}{\langle l_3l_5\rangle\langle l_2l_4\rangle}\right)
\end{align}


The generalized unitarity cuts $A$ and $C$ may seem identical to the generalized unitarity cuts $a$ and $c$ found in section \ref{GU afsnit}. The difference comes from some off-shell momentum flowing into the form factors because of the gauge-invariant operator, and hence the momenta of the on-shell legs no longer sum to zero:

\begin{align}
l_1+l_3+l_4+l_6&\neq0&&\textrm{in Cut}_A,\\
l_1+l_2+l_3+l_4&\neq0&&\textrm{in Cut}_C
\end{align}

We introduce functions $a$ and $b$ as in equation \eqref{4-pt opdeling}. The two functions will be written in terms of the integrals shown in figure \ref{Andet Eksempel Integraler}. The integrals are however not just scalar integrals like in section \ref{GU afsnit}. The integrals ${\rm DBy(1,2|3,4)}$ and ${\rm TriPy(1|2,3,4)}$ include a numerator factor:

\begin{figure}
\begin{center}
\begin{tabular}{ccc}
\includegraphics[scale=0.5]{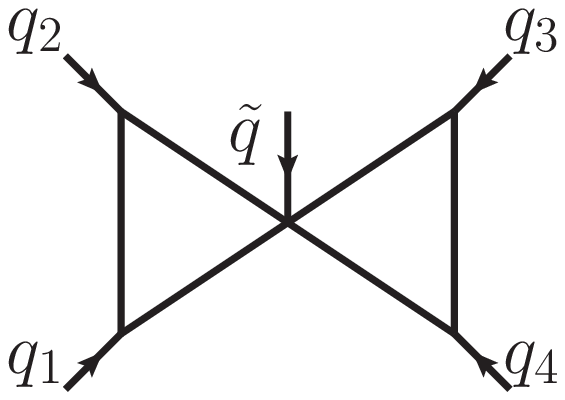}&\includegraphics[scale=0.5]{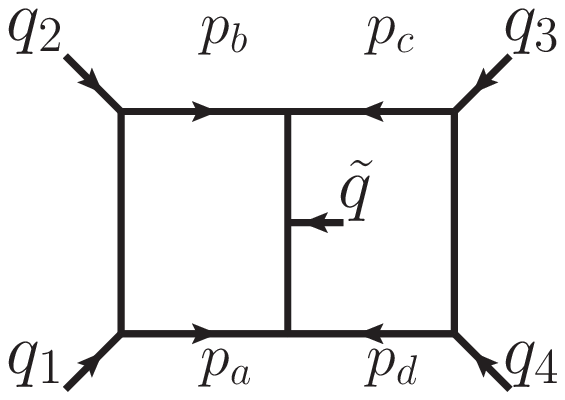}&\includegraphics[scale=0.5]{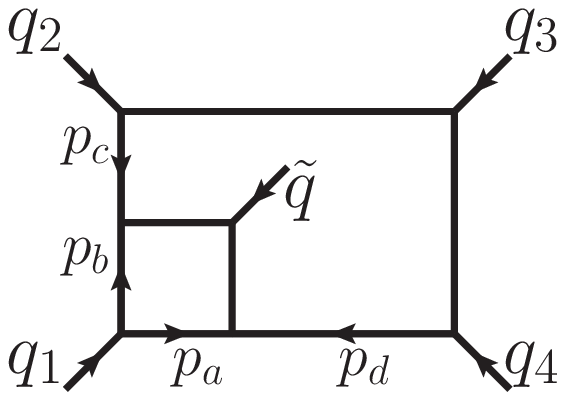}\\
${\rm BTiey(1,2,3,4)}$&${\rm DBy(1,2|3,4)}$&${\rm TriPy(1|2,3,4)}$\\
\includegraphics[scale=0.5]{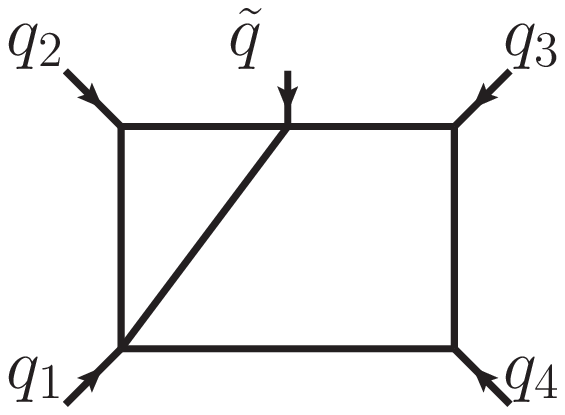}&\includegraphics[scale=0.5]{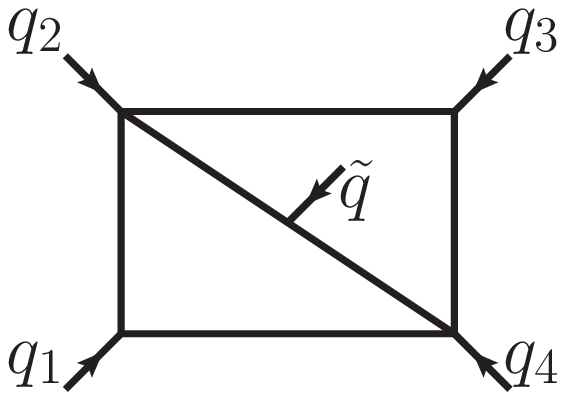}&\includegraphics[scale=0.5]{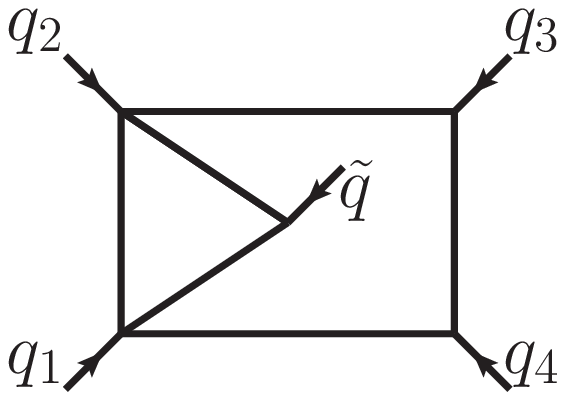}\\
${\rm TriBy(1|2|3,4)}$&${\rm DBy2(1,2|3,4)}$&${\rm TriPy2(1|2|3,4)}$
\end{tabular}
\caption{Integrals used for the 4-point example. The $q_i$'s are the momenta associated with the gauge-invariant operators at the points $x_i$ while $\tilde{q}$ is the momentum associated with the Lagrangian insertion\label{Andet Eksempel Integraler}}
\end{center}
\end{figure}

\begin{align}
{\rm Numerator}=&{}{\rm Sp}(P_+\cancel{p}_a\cancel{p}_b\cancel{p}_c\cancel{p}_d),
\end{align}
\noindent where ${\rm Sp}$ is the trace over spinor indices and $P_+$ is a projector such that if the momenta are on-shell the numerator becomes:

\begin{align}
{\rm Numerator}\Bigg|_{p_a,p_b,p_c,p_d\ {\rm on-shell}}\!\!\!\!\!\!\!\!\!\!\!\!=&{}\langle ab\rangle[bc]\langle cd\rangle[da].
\end{align}

Apart from these two integrals the rest are simple scalar integrals. In terms of these integrals, the functions $a$ and $b$ can be determined from the generalized unitarity cuts above to be:

\begin{align}
\tilde{a}(1,2)=&{}-{\rm BTiey}(1,2,3,4),\label{a-tilde y}\\
\tilde{b}(1,2,3,4)=&{}-{\rm DBy(1,2|3,4)}-{\rm DBy(2,3|4,1)}-{\rm TriPy(1|2,3,4)}-{\rm TriPy(2|3,4,1)}\label{b-tilde y}\\
&-{\rm TriPy(3|4,1,2)}-{\rm TriPy(4|1,2,3)}-{\rm TriBy(1|2|3,4)}-{\rm TriBy(2|3|4,1)}\nonumber\\
&-{\rm TriBy(3|4|1,2)}-{\rm TriBy(4|1|2,3)}-{\rm TriBy(1|4|3,2)}-{\rm TriBy(2|1|4,3)}\nonumber\\
&-{\rm TriBy(3|2|1,4)}-{\rm TriBy(4|3|1,1)}+2{\rm DBy2(1,2|3,4)}+2{\rm DBy2(2,3|4,1)}\nonumber\\
&+2{\rm TriPy2(1|2|3,4)}+2{\rm TriPy2(2|3|4,1)}+2{\rm TriPy2(3|4|1,2)}+2{\rm TriPy2(4|1|2,3)}.\nonumber
\end{align}

These expressions are also consistent with the generalized unitarity cuts shown in figure \ref{AndetEksempel unodvendig} though these generalized unitarity cuts can be avoided using the following arguments. As mentioned previously, operator product expansion arguments \cite{Eden:2011we} lead to the conclusion that the connected diagrams for the relevant correlation functions only contain simple poles, such as:

\begin{figure}
\begin{center}
\begin{tabular}{cc}
\includegraphics[scale=0.8]{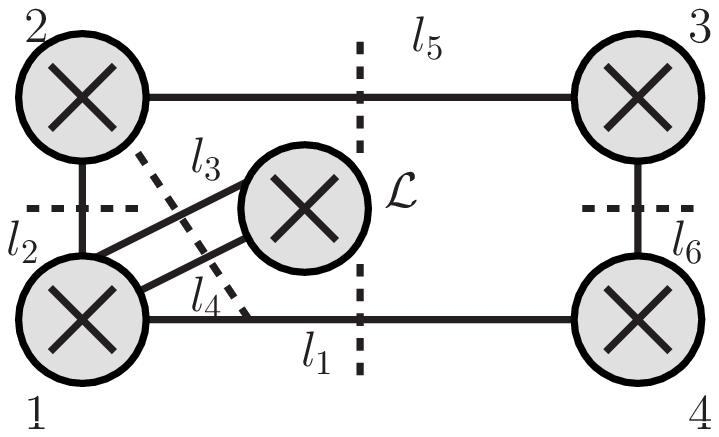}&\includegraphics[scale=0.8]{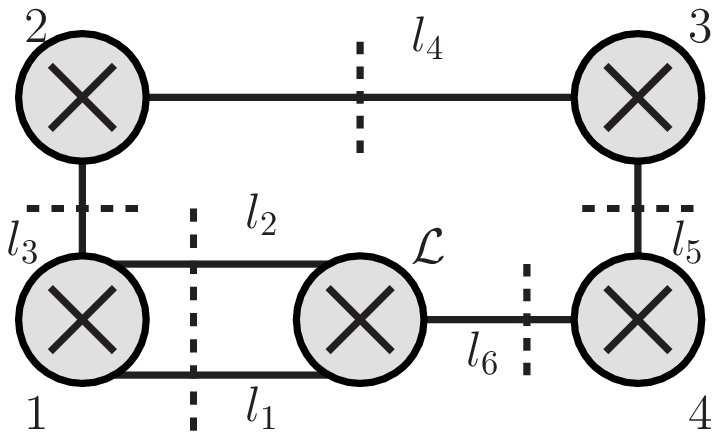}
\end{tabular}
\caption{Necessary generalized unitarity cuts to find the double poles\label{AndetEksempel unodvendig}}
\end{center}
\end{figure}

\begin{align*}
&[(x_i-x_j)^2]^{-1}&&[(x_i-y_j)^2]^{-1}&&[(y_i-y_j)^2]^{-1}.
\end{align*}

For this reason one could ignore the generalized unitarity cuts in figure \ref{AndetEksempel unodvendig} and simply throw away any double poles that appear in the final result.

After Fourier-transforming and introducing the integration over the insertion point, the expressions from \eqref{a-tilde y} and \eqref{b-tilde y} reproduce the results found in section \ref{GU afsnit}. In order to write the functions $a$ and $b$ exactly as in that section, the following relation is useful:

\begin{align}
\frac{\partial}{\partial x_{1}^{[\mu}}\frac{\partial}{\partial x_{2}^{\nu]}}\int\frac{d^4y}{(x_1-y)^2(x_2-y)^2(x_3-y)^2}=&{}-4i\pi^2\frac{(x_1-y)_{[\mu}(x_2-y)_{\nu]}}{(x_1-x_2)^2(x_2-x_3)^2(x_1-x_3)^2}
\end{align}


There are two big differences between this calculation and the one found in section \ref{GU afsnit}. First of all, the single integration variable, $y$, arose naturally as part of the Lagrangian insertion procedure while it came about through a complicated identity in the previous calculation. Secondly, we only needed MHV form factors for this computation while the calculation from section \ref{GU afsnit} required the use of $\overline{\rm MHV}$ form factors. This will become a large advantage at higher loop orders as the previous procedure will require Next-to-MHV quantities, Next-to-next-to-MHV quantities etc. When we apply generalized unitarity in later sections, it will be used after the Lagrangian insertion as done in this section.


\subsection{Scalar Polygon}\label{afsnit skalar}

As reviewed in section \ref{div Wilson}, the scalar polygon will interact like a Wilson loop. We are only interested in the planar theory meaning that the relevant Feynman diagrams or generalized unitarity cuts can all be drawn on a two-dimensional surface. So even though there are more than two space-time dimensions, the diagrams are essentially two-dimensional, and it is meaningful to divide the diagrams into two parts: one inside and one outside the polygon. This explains the origin of the appearance of the amplitude squared in \eqref{duality eq}: the inside of the polygon will give one factor of the amplitude and the outside another.

Our goal will be to show that cuts with all Lagrangians inside the polygon correspond to the generalized unitarity cuts of the corresponding amplitude. The generalization to cuts with Lagrangian insertions both inside and outside will then be straightforward.

\begin{figure}
\begin{center}
\begin{tabular}{cc}
\includegraphics[scale=1]{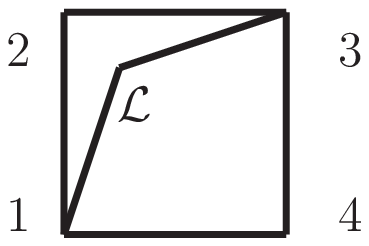}&\includegraphics[scale=1]{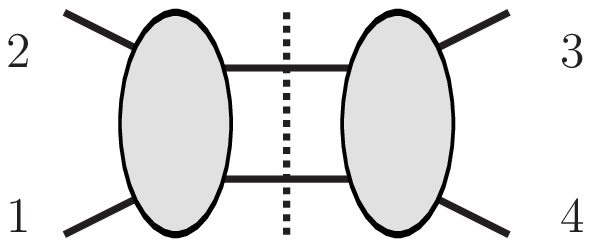}\\
(a)&(b)
\end{tabular}
\end{center}
\caption{A cut of a four-sided polygon and its corresponding generalized unitarity cut\label{cut versus gu-cut}}
\end{figure}

It is important that the cuts separate the inside of the polygon into parts that do not interact except through the shared internal lines. As an example consider the cut in figure \ref{cut versus gu-cut}(a) where the lines represents distances that have been made light-like\footnote{We will be more specific about what we mean by these diagrams later}. This cut will correspond to the generalized unitarity cut in figure \ref{cut versus gu-cut}(b), so there should not be any direct interaction between the sides $x_2-x_3$ and $x_3-x_4$\footnote{Except of course through the outside of the polygon but as mentioned this will be interpreted as part of the other amplitude in the duality}, just like there are no explicit factors of $\langle23\rangle$ or $[23]$ in the generalized unitarity cut.

\begin{figure}
\begin{center}
\begin{tabular}{cc}
\includegraphics[scale=1]{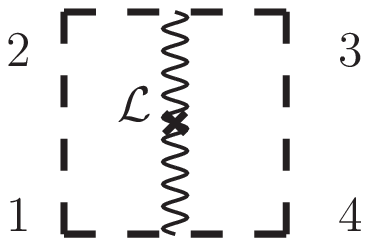}&\includegraphics[scale=1]{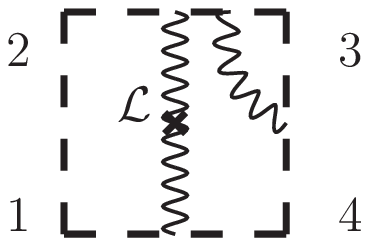}\\
(a)&(b)
\end{tabular}
\end{center}
\caption{An example of a diagram that should contribute to the mentioned cut and a diagram that should not. Dashed lines represent scalars and wiggly lines represent gluons\label{gode og daarlige}}
\end{figure}

It is not immediately obvious that this requirement is satisfied. For instance, the diagram in \ref{gode og daarlige}(a), where a scalar polygon interacts through gluons with a single Lagrangian insertion, will contribute to the cut. However, the diagram in figure \ref{gode og daarlige}(b), which is the same but with an additional gluonic interaction between the two sides of the polygon, would ruin this property and so should not contribute to the cut.

To understand the separation of the polygon, let us consider a side of the polygon spanned between the points $x_i$ and $x_{i+1}$. Let the side be connected through $m$ vertices to $m$ different Lagrangian insertions as shown in figure \ref{separation}. To more easily distinguish between the insertion points and the points on the polygon we will use tildes when enumerating the insertion points and their spinors, harmonic variables and fermionic variables. In accordance with equations \eqref{gammelt1} and \eqref{gammelt2}, we write the scalar line as a regular light-like Wilson line. This means that the diagram will be proportional to $m$ propagators each connecting a point on the Wilson with a Lagrangian insertion point. Each Lagrangian insertion will supply a single derivative so the diagram will be proportional to:

\begin{figure}
\begin{center}
\includegraphics[scale=1]{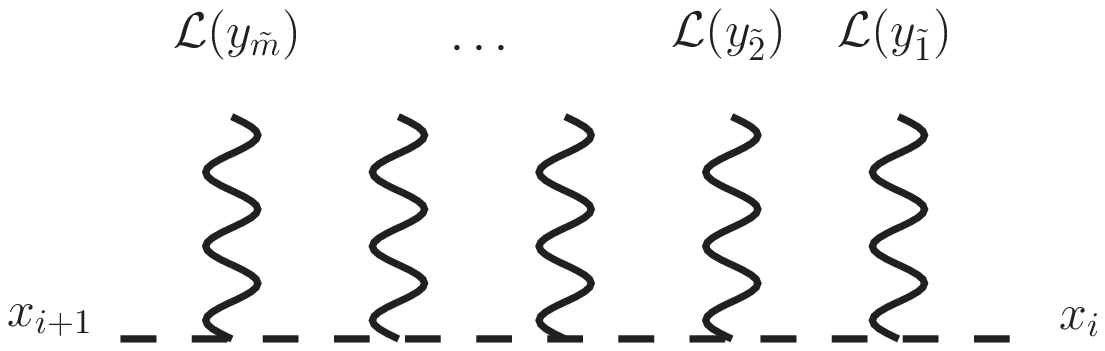}
\end{center}
\caption{One side of a scalar polygon interacting with $m$ Lagrangian insertions\label{separation}}
\end{figure}

\begin{align}
&I_m(y_{\tilde{m}},\cdots y_{\tilde{1}};t_{m+1}=0)\label{string vertices}\\
=&{}\int_0^1dt_m(x_i-x_{i+1})_{[\mu_m}\frac{\partial}{\partial y_{\tilde{m}}^{\nu_m]}}\Delta(y_{\tilde{m}},t_m)\cdots
\int_{t_2}^1dt_1(x_i-x_{i+1})_{[\mu_1}\frac{\partial}{\partial y_{\tilde{1}}^{\nu_1]}}\Delta(y_{\tilde{1}},t_1),\nonumber
\end{align}
\noindent where the propagators are given by:

\begin{align}
\Delta(y_{\tilde{\jmath}},t_j)=&{}\frac{1}{(x_{i+1}-y_{\tilde{\jmath}})^2(1-t_j)+(x_i-y_{\tilde{\jmath}})^2t_j}.
\end{align}

\begin{figure}
\begin{center}
\includegraphics[scale=1]{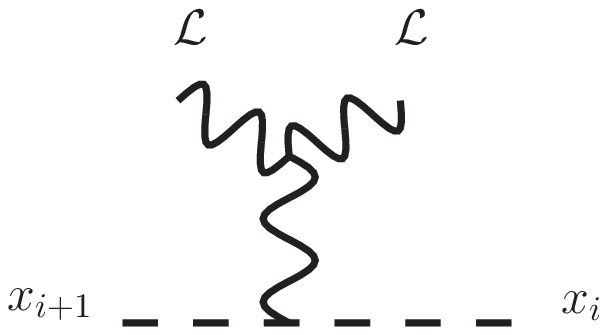}
\end{center}
\caption{One side of a scalar polygon interacting with $m$ Lagrangian insertions\label{separation3}}
\end{figure}

The integral is relevant to more than just the case shown in figure \ref{separation}. For instance, we may add gluon vertices as exemplified in figure \ref{separation3}. From the arguments in section \ref{div Wilson}, we see that the derivative from the vertex must the counter the effect of the additional propagator. Since there is only one derivative in the gluon vertex, the divergence can only be upheld for one of the two Lagrangians. Consequently, the cases with added gluon vertices will have the same divergence behaviour as in \eqref{string vertices}.

Let us proceed to study the behaviour of \eqref{string vertices} when the insertion points become light-like separated from point on the polygon. In the following there will be a caveat relating to cases where a single insertion point become light-like separated from both points on the polygon. This particular case will be dealt with at the end of the section. We begin  by studying the right-most integral in \eqref{string vertices}:

\begin{align}
\int_{t_2}^1dt_1(x_i-x_{i+1})_{[\mu_1}\frac{\partial}{\partial y_{\tilde{1}}^{\nu_1]}}\Delta(y_{\tilde{1}},t_1)=&{}\frac{2(x_{i+1}-y_{\tilde{1}})_{[\mu_1}(x_i-x_{i+1})_{\nu_1]}(1-t_2)}{(x_i-y_{\tilde{1}})^2\big[(x_{i+1}-y_{\tilde{1}})^2(1-t_2)+(x_i-y_{\tilde{1}})^2t_2\big]}\label{right-most integral}
\end{align}

This clearly becomes divergent when the distance between $x_i$ and $y_{\tilde{1}}$ become light-like. From the point of view of the integral, this divergence arises because the integrand becomes proportional to $(1-t_1)^{-1}$ which diverges in the upper limit. Notice also that if $x_i$ and $y_{\tilde{1}}$ are not light-like separated, \eqref{right-most integral} contributes with a factor of $(1-t_2)$ which would ruin the divergences for the subsequent Lagrangian. Indeed, it should ruin the divergence for all subsequent Lagrangians since the addition of one propagator and one derivative should not raise the divergence in accordance with the arguments in section \ref{div Wilson}. If $x_i$ and $y_{\tilde{1}}$ are light-like separated the integral will not influence the subsequent integrals and one can do the same analysis for the second right-most integral\footnote{The observant reader will notice that one could also make $y_{\tilde{1}}$ light-like separated from both $x_i$ and $x_{i+1}$ and not worry about the remaining Lagrangian insertions. When including the spinor structure of the on-shell Lagrangian the special three-point kinematics makes the spinors $\lambda^\alpha$ for the three sides of the light-like triangle proportional to each other. This type of limits though interesting will not be relevant to our analysis but would be important if considering maximal cuts}. This argument suggests that integrals of the type in \eqref{string vertices} satisfy the relation:

\begin{align}
\lim_{(x_i-y_{\tilde{j}})^2=0}\frac{I_m(y_{\tilde{m}},\cdots y_{\tilde{1}};t_{m+1})}{\Delta(y_{\tilde{1}},1)}=&{}\Upsilon_{\mu_1\nu_1}(y_{\tilde{1}})\Delta(y_{\tilde{1}},0)I_{m-1}(y_{\tilde{m}},\cdots y_{\tilde{2}};t_{m+1}).\label{graense}
\end{align}
\noindent where the following quantity has been defined:
\begin{align}
\Upsilon_{\mu_j\nu_j}(y_{\tilde{\jmath}})=&{}2(x_{i+1}-y_{\tilde{\jmath}})_{[\mu_j}(x_i-x_{i+1})_{\nu_j]}.
\end{align}

In appendix \ref{integral app}, the integrals have been computed up to $m=4$, and they do satisfy this relation. Note that $I_m$ have some logarithmic divergences that are being removed by the limit \eqref{graense}. We can ignore these terms since the integrand should not contain such divergences at the Born level as mentioned in section \ref{LI afsnit}.


Equation \eqref{graense} can be divided into a part dependent on $y_{\tilde{1}}$ and an integral independent of $y_{\tilde{1}}$. The integral is exactly the same type as the original integral, only with one less insertion point. The above arguments can then be applied to $y_{\tilde{2}}$. Setting $(x_i-y_{\tilde{2}})^2=0$ will give a part dependent on $y_{\tilde{2}}$ and an integral independent of $y_{\tilde{2}}$. The integral will be of the same type as original, and the arguments can then be repeated for $y_{\tilde{3}}$ and so forth.

Consequently we find that the diagram in figure \ref{separation} only contributes to the cut where a specific $y_{\tilde{\jmath}}$ becomes light-like separated from $x_i$ if all the Lagrangians to the right of $y_{\tilde{\jmath}}$ are also light-like separated from that point. Similarly, the diagram only contributes to the cut where $y_{\tilde{\jmath}}$ becomes light-like separated from $x_{i+1}$ if all the Lagrangians to the left of $y_{\tilde{\jmath}}$ are also light-like separated from that point. This shows that the necessary separation does appear.

Making an insertion point $y_{\tilde{\jmath}}$ light-like separated from $x_i$ lead to the following factor dependent on the insertion point:

\begin{align*}
\Upsilon_{\mu_j\nu_j}(y_{\tilde{\jmath}})\Delta(y_{\tilde{\jmath}},0).
\end{align*}

If we include the spinor structure from the on-shell Lagrangian and define the spinors $\lambda_{\tilde{\jmath}}^\alpha\tilde{\lambda}^{\dot{\alpha}}_{\tilde{\jmath}}=(x_i-y_{\tilde{\jmath}})^{\alpha\dot{\alpha}}$, this gives the following quantity:

\begin{align}
\tfrac{1}{2}\epsilon^{\dot{\beta}\dot{\alpha}}(\sigma^{\mu_j})_{\alpha\dot{\alpha}}(\sigma^{\nu_j})_{\beta\dot{\beta}}\Upsilon_{\mu_j\nu_j}(y_{\tilde{\jmath}})\Delta(y_{\tilde{\jmath}},0)=&{}-\frac{\lambda_{\tilde{\jmath}(\alpha}\lambda_{i\beta)}}{\langle i\tilde{\jmath}\rangle}.\label{equation1}
\end{align}

This contribution would then have to be added to the one with the Wilson loop vertex on the other side of $x_i$ which can be found through an equivalent calculation though the sign will be opposite\footnote{The polygon interacts like two Wilson loops with opposite directions, it is the direction that introduces this sign. It is arbitrary which of the two Wilson loops we choose to consider}. Focusing on $y_{\tilde{1}}$ for the moment, the light-like gives us:

\begin{align}
-\frac{\lambda_{\tilde{1}(\alpha}\lambda_{i\beta)}}{\langle i\tilde{1}\rangle}+\frac{\lambda_{\tilde{1}(\alpha}\lambda_{i-1\beta)}}{\langle i{-}1\tilde{1}\rangle}&=\frac{\langle i-1i\rangle}{\langle i-1\tilde{1}\rangle\langle \tilde{1}i\rangle}\lambda_{\tilde{1}(\alpha}\lambda_{\tilde{1}\beta)}\label{equation2}.
\end{align}

Additional vertices can be added on the gluon line connecting the scalar polygon with $y_{\tilde{1}}$, and one can show that they will act like the Wilson line vertices. This point is slightly non-trivial as the counting arguments from \cite{Engelund:2011fg} do not remove all of the unwanted terms.

\begin{figure}
\begin{center}
\includegraphics[scale=0.7]{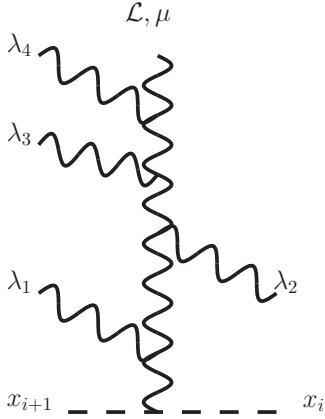}
\end{center}
\caption{An example with added three-gluon vertices. Here $r=4$\label{ekstra Wilson}}
\end{figure}

Consider diagrams with $r$ three-gluon vertices like the one shown in figure \ref{ekstra Wilson}\footnote{The arguments are presented in Feynman gauge but it is simple to extend the arguments to more general gauges}. In the light-like limit each three-gluon vertex will contribute with a vector\footnote{There will also be vectors $(x_{i+1}-y_{\tilde{1}})^{\kappa}$ but they will come with a factor of $1-t$. As in the argument leading to \eqref{graense}, we will discard these terms}:

\begin{align}
(x_i-y_{\tilde{1}})^{\kappa}.
\end{align}

In addition to these $r$ vectors, there is the vector coming from the scalar line:

\begin{align}
(x_i-x_{i+1})^\kappa.
\end{align}

There are also $r+1$ Lorentz indices: one for each outgoing gluon (the $\lambda$'s in figure \ref{ekstra Wilson}) and the one free index from the field strength from the inserted Lagrangian (the $\mu$ in the figure).

Each vector can be assigned one of the $r+1$ Lorentz indices, or their Lorentz indices can be contracted by introducing a metric tensor. Antisymmetry removes any terms where $\mu$ is assigned to a vector $(x_i-y_{\tilde{1}})$. Because of the light-like limit the terms proportional to $(x_i-y_{\tilde{1}})^2$ also go away. The option, where one of the vectors from the three-gluon vertices is multiplied the vector from the scalar line, gives something not dependent on $x_{i+1}$. This is because the product of the two vectors removes the propagator factor:

\begin{align}
2(x_i-y_{\tilde{1}})\cdot(x_i-x_{i+1})=&{}-(x_{i+1}-y_{\tilde{1}})^2.
\end{align}

These terms then cancel against the similar terms from the other side of $x_i$ where the factor $(x_{i-1}-y_{\tilde{1}})^2$ has been removed. The remaining term behaves as if the extra vertices where Wilson line vertices on the light-like line from $x_i$ to $y_{\tilde{1}}$.

As reviewed in section \ref{div Wilson}, the light-like Wilson line act identical to a scalar propagator between two light-like separated points. If we therefore replace the bilinear scalar operator at $x_i$ by a cubic operator and the on-shell Lagrangian by an operator proportional to $\mathcal{Tr}(F^{\alpha\beta}\phi^{++})$, it should behave in the same way. $\mathcal{Tr}(F^{\alpha\beta}\phi^{++})$ in fact appears in the chiral part of the stress-tensor multiplet. Diagrammatically the relation can put in the form:

\begin{align}
\lim_{(x_i-y_{\tilde{1}})^2=0}\frac{(x_i-y_{\tilde{1}})^2}{\bf (i\tilde{1})}&\int d^4\theta_{\tilde{1}}\left.\begin{minipage}{2.5cm}
\includegraphics[scale=0.8]{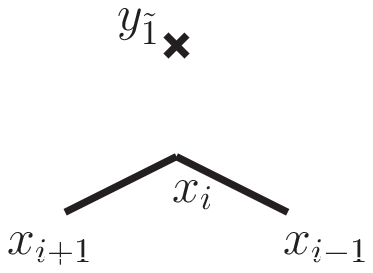}
\end{minipage}\qquad\right|_{\theta_{i-1}^+=\theta_i^+=\theta_{i+1}^+=0}\label{vertex relation}\\
&=\frac{1}{{\bf (i\tilde{1})}}\frac{\langle i-1i\rangle}{\langle i-1\tilde{1}\rangle\langle \tilde{1}i\rangle}\int d^4\theta_{\tilde{1}}\delta^2(\langle \tilde{1}\theta_{\tilde{1}}^{+a}\rangle)\left.\begin{minipage}{2.5cm}
\includegraphics[scale=0.8]{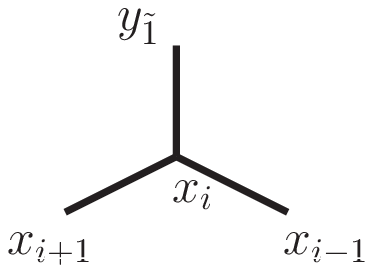}\nonumber
\end{minipage}\qquad\right|_{\theta_{i-1}^+=\theta_i^+=\theta_{i+1}^+=0},
\end{align}
\noindent where full lines represent distances made light-like after dividing by a scalar propagator, and vertices where $d$ lines meet correspond to local operators of the type $\mathcal{T}_d$ (the operators at $x_{i-1}$, $x_{i+1}$ and $y_{\tilde{1}}$ are connected to other operators not represented in the diagrams, and we have suppressed a numerical factor including the coupling constant)\footnote{A version of this relation also appeared in \cite{CaronHuot:2010ek} where it was used to establish a BCFW relation at loop level}.

Because the line connecting $x_i$ and $y_{\tilde{1}}$ acts like a regular Wilson line, it is straightforward to generalize this. Making $y_{\tilde{2}}$ light-like separated from $x_i$ gives a factor similar to \eqref{equation2}, only now with $y_{\tilde{1}}$ playing the role of $x_{i-1}$:

\begin{align}
-\frac{\lambda_{\tilde{2}(\alpha}\lambda_{i\beta)}}{\langle i\tilde{2}\rangle}+\frac{\lambda_{\tilde{2}(\alpha}\lambda_{\tilde{1}\beta)}}{\langle \tilde{1}\tilde{2}\rangle}&=\frac{\langle \tilde{1}i\rangle}{\langle \tilde{1}\tilde{2}\rangle\langle \tilde{2}i\rangle}\lambda_{\tilde{2}(\alpha}\lambda_{\tilde{2}\beta)}
\end{align}

The arguments concerning $y_{\tilde{1}}$ can then be repeated for $y_{\tilde{2}}$. The diagrams will behave as if there was a light-like Wilson line between $x_i$ and $y_{\tilde{2}}$. This can be described by replacing the operator at $x_i$ by a quartic scalar operator and the operator at $y_{\tilde{2}}$ by one proportional to $\mathcal{Tr}(F^{\alpha\beta}\phi^{++})$.

Generalizing to more insertion is then straightforward. Generalizing to the supersymmetric case requires something more effective than the Feynman diagram approach used here. In the next section we will use generalized unitarity cuts to discuss the supersymmetrization of \eqref{vertex relation}.

Before this, we will however return to the case of and insertion point becoming light-like separated from both $x_i$ and $x_{i+1}$. Equation \eqref{right-most integral} shows  that if $(x_i-y_{\tilde{1}})^2$ is set to 0, the integral will also have a simple pole in $(x_{i+1}-y_{\tilde{1}})^2$ regardless of how many other Lagrangian insertions appears to the left of $y_{\tilde{1}}$ in figure \ref{separation3}. However this is a very special situation as it requires that:

\begin{align}
[\tilde{1}i]\langle i\tilde{1}\rangle=&{}0.
\end{align}

This means that either $\lambda_i$ is proportional to $\lambda_{\tilde{1}}$ or $\tilde{\lambda}_i$ is proportional to $\tilde{\lambda}_{\tilde{1}}$. Using the spinor structure from the on-shell Lagrangian as in equation \eqref{equation1}, we get:

\begin{align}
\lim_{(x_{i+1}-y_{\tilde{1}})^2=0}(x_{i+1}-y_{\tilde{1}})^2\tfrac{1}{2}\epsilon^{\dot{\beta}\dot{\alpha}}(\sigma^{\mu_1})_{\alpha\dot{\alpha}}(\sigma^{\nu_1})_{\beta\dot{\beta}}\Upsilon_{\mu_1\nu_1}(y_{\tilde{1}})\Delta(y_{\tilde{1}},0)=&{}\lambda_{\tilde{1}(\alpha}\lambda_{i\beta)}[ i\tilde{1}]
\end{align}

For this to be non-zero, it must be $\lambda_i$ which is proportional to $\lambda_{\tilde{1}}$. The amplitudes interpretation of these types of limits would be generalized unitarity cuts involving 3-point $\overline{\rm MHV}$ amplitudes. Though such generalized unitarity cuts can be very useful, they are not necessary to describe the loop amplitudes in $\mathcal{N}=4$ super Yang-Mills. Nothing forces to consider the limits for the correlation functions either, and the subsequent sections it will be easier to avoid them.

\subsection{Supersymmetrization}\label{susy afsnit}

In order to find the correct supersymmetrization of the cuts, we are going to use generalized unitarity. As described in section \ref{GU on LI}, we will first use Lagrangian insertion to give us a Born-level correlator then consider the generalized unitarity cuts of that correlator. As mentioned in that section, this makes the generalized unitarity be made up entirely of MHV form factors which can be found in section \ref{GU afsnit}.

We are not going to compute the full generalized unitarity cuts only draw certain conclusions about the fermionic structure of the position space cuts. We will only consider the generalized unitarity cuts where the operators made light-like separated are connected through a cut propagator. This is sufficient as long as we avoid the limits described at the end of the previous section where $\lambda$-spinors become proportional to each other.

To see that these generalized unitarity cuts are indeed sufficient, consider the following. For there to be a divergence when two operators are made light-like separated, they must be connected through some sequence of propagators and vertices. As explained under equation \eqref{cut imellem op} those propagators cannot all be canceled because that would lead to a delta function, so there must be some propagator that can be cut for the terms to contribute to the light-like limit. It is for this reason we only consider generalized unitarity cuts where each light-like distance has a corresponding cut propagator.

As an example consider the 4-point function again. Assume we are interested in the position space cut where the following distances are made light-like:

\begin{align}
(x_2-y)^2=(x_4-y)^2=(x_1-x_2)^2=(x_2-x_3)^2=(x_3-x_4)^2=(x_1-x_4)^2=&{}0.
\end{align}

To study this position space cut we need only consider one generalized unitarity cut, namely ${\rm Cut}_D$. Not that the other generalized unitarity cuts in figure \ref{AndetEksempel cuts} do not capture terms relevant to this limit, they do. In fact ${\rm Cut}_A$, ${\rm Cut}_B$ and ${\rm Cut}_E$ all include terms that survive in this particular limit. However they are not guaranteed to capture all terms relevant to the position space cut, and the additional information stored in those generalized unitarity cuts relates to parts of the correlation function that are removed in the above limit. On the other hand, ${\rm Cut}_D$ capture all terms relevant in this limit. It is therefore sufficient to study ${\rm Cut}_D$ in order to learn about the particular position space cut described above.

This means that it is sufficient when dealing with the super-correlators/super-amplitudes duality to study generalized unitarity cuts where the form factors create the same polygon as used for the duality ($i.e.$ we will be interested in generalized unitarity cuts like ${\rm Cut}_D$ and ${\rm Cut}_E$ where the form factor for the operator at the point $x_1$ is connected to the form factor for the operator at the point $x_2$, and the form factor for the operator at point $x_2$ is connected to the form factor for the operator at the point $x_3$ etc.).

As a consequence, it will still make sense to divide the planar diagrams into a part inside and a part outside of the polygon. For a correlation function $G_n$ there will be $n$ cut propagators connecting the form factors associated with the operators at the original points on the polygon. Each of the form factors will contribute with 8 fermionic delta functions which means there will be $8n$ fermionic delta functions depending on the aforementioned $n$ cut propagators. After performing the $4n$ Grassmann integrations associated with the cut propagators, we will be left with $4n$ fermionic delta functions all depending on spinor products where one of the spinors correspond to momentum flowing along a side of the polygon. In the light-like limit the spinor products will either cancel similar spinor products in the denominator or be part of derivatives becoming proportional to the position space spinors \eqref{real space spinors}. Consequently those fermionic delta functions will correspond to either the outside or the inside of the polygon interacting with the sides of the polygon. There will be no direct interactions between the inside and the outside of the polygon. Planarity ensures that the denominators on the polygon as well as factors not part of the polygon will not give such direct interactions either.

Let us proceed to generalize \eqref{vertex relation}. We are going to start with an ansatz and use generalized unitarity to confirm it. Our ansatz will be that the two fermionic delta functions get replaced by\footnote{To avoid confusing with $\sqrt{-1}$ we replace the index $i$ with the index $j$ up until \eqref{vertex relation susy}}:

\begin{align}
&\delta^2(\chi_{\tilde{1}/\tilde{1}}^a-\langle \tilde{1}\theta_j^A\rangle(\tilde{1})_A^{+a}),\label{deltafunktion}
\end{align}
\noindent where $(\tilde{1})_A^{+a}$ are the harmonic variables associated with the Lagrangian insertion at $y_{\tilde{1}}$. We use $\chi_{\tilde{1}/\tilde{1}}^a$ to denote $\langle \tilde{1}\theta_{\tilde{1}}^{+a}\rangle$ similar to the notation in \eqref{fermionic identification}. One should note that $\theta_{j\alpha}^A$ does not appear freely in the construction of the correlation functions as that would give twice as many Grassmann variables as for the scattering amplitudes. It only appears as part of very specific products with spinors and harmonic variables so the second term should be interpreted in terms of the following:

\begin{align}
\langle jj-1\rangle\langle \tilde{1}\theta_j^A\rangle=&{}\langle\tilde{1}j-1\rangle\langle j\theta_{j+1}^{+a}\rangle (\overline{\jmath} j+1)^{-1\phantom{a}a'}_{\phantom{-1}a}(\overline{\jmath})_{-a'}^A
+\langle\tilde{1}j-1\rangle\langle j\theta_{j}^{+a}\rangle (\overline{\jmath+1} i)^{-1\phantom{a}a'}_{\phantom{-1}a}(\overline{\jmath+1})_{-a'}^A\label{interpret}\\
&+\langle j\tilde{1}\rangle\langle j-1\theta_j^{+a}\rangle(\overline{\jmath-1} i)^{-1\phantom{a}a'}_{\phantom{-1}a}(\overline{\jmath-1})_{-a'}^A
+\langle j\tilde{1}\rangle\langle j-1\theta_{j-1}^{+a}\rangle(\overline{\jmath} j-1)^{-1\phantom{a}a'}_{\phantom{-1}a}(\overline{\jmath})_{-a'}^A\nonumber.
\end{align}

The factors $(\overline{\imath}j)^{-1\phantom{a}a'}_{\phantom{-1}a}$ are the inverse matrices of $(\bar{\imath})^A_{-a'}(j)^{+a}_A$. In order to find the effect of this delta function on the super-Fourier transform of the form factors, we write it in terms of an integral:

\begin{align}
\delta^2(\chi_{\tilde{1}/\tilde{1}}^a-\langle \tilde{1}\theta_j^A\rangle(\tilde{1})_A^{+a})=&{}-\frac{1}{\langle jj-1\rangle}\int d^2\gamma e^{i\langle jj-1\rangle\langle\tilde{1}\theta_{\tilde{1}}^{+a}\rangle\gamma_a}e^{-i\langle\tilde{1}j-1\rangle\langle j\theta_{j+1}^{+b}\rangle (\overline{\jmath} j+1)^{-1\phantom{b}b'}_{\phantom{-1}b}(\overline{\jmath})_{-b'}^A(\tilde{1})_A^{+a}\gamma_a}\nonumber\\
&e^{-i\big(\langle\tilde{1}j-1\rangle\langle j\theta_{j}^{+b}\rangle (\overline{\jmath+1} i)^{-1\phantom{b}b'}_{\phantom{-1}b}(\overline{\jmath+1})_{-b'}^A+\langle j\tilde{1}\rangle\langle j-1\theta_j^{+b}\rangle(\overline{\jmath-1} j)^{-1\phantom{b}b'}_{\phantom{-1}b}(\overline{\jmath-1})_{-b'}^A\big)(\tilde{1})_A^{+a}\gamma_a}\\
&e^{-i\langle j\tilde{1}\rangle\langle j-1\theta_{j-1}^{+b}\rangle(\overline{\jmath} j-1)^{-1\phantom{b}b'}_{\phantom{-1}b}(\overline{\jmath})_{-b'}^A(\tilde{1})_A^{+a}\gamma_a}\nonumber.
\end{align}

When multiplied by the form factors, these exponents can be removed by shifting the fermionic integration variables for the form factors as follows:

\begin{align}
\hat{\gamma}_{\tilde{1}+a}^\alpha&=\gamma_{\tilde{1}+a}^\alpha-\langle jj-1\rangle\lambda_{\tilde{1}}^\alpha\gamma_a\nonumber,\\
\hat{\gamma}_{j-1+a}^\alpha&=\gamma_{j-1+a}^\alpha+\langle j\tilde{1}\rangle\lambda_{j-1}^\alpha(\overline{\jmath} j-1)^{-1\phantom{a}a'}_{\phantom{-1}a}(\overline{\jmath})_{-a'}^A(\tilde{1})_A^{+b}\gamma_a,\label{shifts}\\
\hat{\gamma}_{j+a}^\alpha&=\gamma_{j+a}^\alpha+\big(\langle\tilde{1}j-1\rangle\lambda_j^\alpha (\overline{\jmath+1} j)^{-1\phantom{a}a'}_{\phantom{-1}a}(\overline{\jmath+1})_{-a'}^A+\langle j\tilde{1}\rangle\lambda_{j-1}^\alpha(\overline{\jmath-1} j)^{-1\phantom{a}a'}_{\phantom{-1}a}(\overline{\jmath-1})_{-a'}^A\big)(\tilde{1})_A^{+b}\gamma_b\nonumber,\\
\hat{\gamma}_{j+1+a}^\alpha&=\gamma_{j+1+a}^\alpha+\langle \tilde{1}j-1\rangle\lambda_{j}^\alpha(\overline{\jmath} j+1)^{-1\phantom{a}a'}_{\phantom{-1}a}(\overline{\jmath})_{-a'}^A(\tilde{1})_A^{+b}\gamma_b\nonumber.
\end{align}

The delta function \eqref{deltafunktion} is, thereby, replaced by imposing the invariance under a specific shift of the fermionic variables. We may write this as:

\begin{align}
&\int d^2\gamma d^4\gamma_{\tilde{1}}d^4\gamma_{j-1}d^4\gamma_jd^4\gamma_{j+1}\delta^2(\chi_{\tilde{1}/\tilde{1}}^a-\langle \tilde{1}\theta_j^A\rangle(\tilde{1})_A^{+a})e^{i\theta_{\tilde{1}\alpha}^{+a}\gamma_{\tilde{1}+a}^\alpha+i\theta_{j-1\alpha}^{+1}\gamma_{j-1+a}^\alpha+i\theta_{j\alpha}^{+1}\gamma_{j+a}^\alpha+i\theta_{j+1\alpha}^{+1}\gamma_{j+1+a}^\alpha}\nonumber\\
&\mathcal{F}\big(\gamma_{\tilde{1}+a}^\alpha,\gamma_{j-1+a}^\alpha,\gamma_{j+a}^\alpha,\gamma_{j+1+a}^\alpha,\cdots\big)\label{fermi1}\\
&\qquad\qquad\qquad=-\frac{1}{\langle jj-1\rangle}\int d^2\gamma d^4\hat{\gamma}_{\tilde{1}}d^4\hat{\gamma}_{j-1}d^4\hat{\gamma}_jd^4\hat{\gamma}_{j+1}e^{i\theta_{\tilde{1}\alpha}^{+a}\hat{\gamma}_{\tilde{1}+a}^\alpha+i\theta_{j-1\alpha}^{+1}\hat{\gamma}_{j-1+a}^\alpha+i\theta_{j\alpha}^{+1}\hat{\gamma}_{j+a}^\alpha+i\theta_{j+1\alpha}^{+1}\hat{\gamma}_{j+1+a}^\alpha}\nonumber\\
&\qquad\qquad\qquad\phantom{=}\mathcal{F}\big(\gamma_{\tilde{1}+a}^\alpha,\gamma_{j-1+a}^\alpha,\gamma_{j+a}^\alpha,\gamma_{j+1+a}^\alpha,\cdots\big)\nonumber,
\end{align}
\noindent where the $\gamma^\alpha$'s are now functions of the $\hat{\gamma}^\alpha$'s and $\gamma_a$. 

\begin{figure}
\begin{center}
\includegraphics[scale=0.8]{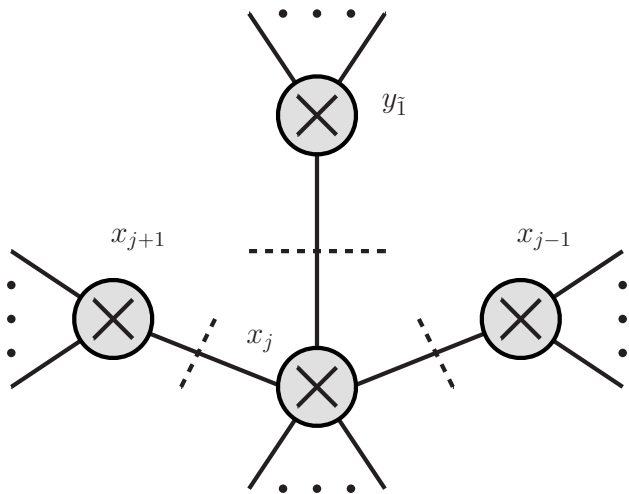}
\caption{Part of generalized unitarity cut\label{shift figur}}
\end{center}
\end{figure}

The effect of this shift symmetry can be seen by studying the part of the generalized unitarity cuts shown in figure \ref{shift figur}. An MHV form factor associated with an operator $\mathcal{T}_d$ placed at $x_j$ is connected to form factors for operators at $x_{j-1}$, $x_{j+1}$ and $y_{\tilde{1}}$. This section will be connected to the rest of the generalized unitarity which will differ from case to case. However these elements will be present in all the generalized unitarity cuts necessary to capture the behaviour in the limit where:

\begin{align}
(x_j-x_{j-1})^2=(x_j-x_{j+1})^2=(x_j-y_{\tilde{1}})^2&=0.
\end{align}

The momenta of the on-shell legs connecting the form factors are denoted $P_{\tilde{1}}$, $P_{j-1}$ and $P_j$. Since these momenta are responsible for the light-like divergences, we may replace the position space spinors $\lambda_{\tilde{1}}$, $\lambda_{j-1}$ and $\lambda_{j}$ by the momentum spinors $\lambda_{P_{\tilde{1}}}$, $\lambda_{P_{j-1}}$ and $\lambda_{P_j}$. This can be seen from inverting the arguments found in section \ref{div Wilson}. There the momenta were all replaced by light-like vectors. Here we replace the light-like vectors with momenta. The cost of replacing position space spinors with momentum spinors consists of a rescaling $\gamma_a$ and a bosonic factor. Since the purpose is only to show that the ansatz for the fermionic part \eqref{deltafunktion} is correct, we will not be interested in bosonic factors.

The relevant quantity is then the super-momentum conserving delta functions from the four form factors in the figure and $\widetilde{\mathcal{F}}_{\mathcal{T}_d}$ for the operator at point $x_j$:

\begin{align}
&\int d^4\eta_{P_{j-1}}d^4\eta_{P_{j}}d^4\eta_{P_{\tilde{1}}}\int d^2\gamma \delta^8\left(\gamma_{j-1+a}^\alpha(j-1)^{+a}_A-\sum_r\eta_{rA}\lambda_r^\alpha-\eta_{P_{j-1}A}\lambda_{P_{j-1}}^\alpha\right)\nonumber\\
&\delta^8\left(\gamma_{j+1+a}^\alpha(j+1)^{+a}_A-\sum_r\eta_{rA}\lambda_r^\alpha-\eta_{P_{j}A}\lambda_{P_{j}}^\alpha\right)\delta^8\left(\gamma_{\tilde{1}+a}^\alpha(\tilde{1})^{+a}_A-\sum_r\eta_{rA}\lambda_r^\alpha-\eta_{P_{\tilde{1}}A}\lambda_{P_{\tilde{1}}}^\alpha\right)\label{fermi2}\\
&\delta^8\left(\gamma_{j+a}^\alpha(j)^{+a}_A-\sum_r\eta_{rA}\lambda^\alpha_r+\eta_{P_{j-1}A}\lambda_{P_{j-1}}^\alpha+\eta_{P_{j}A}\lambda_{P_{j}}^\alpha+\eta_{P_{\tilde{1}}A}\lambda_{P_{\tilde{1}}}^\alpha\right)\nonumber\\
&\widetilde{\mathcal{F}}^{MHV}_{\mathcal{T}_d}\big(\gamma_{j+a}^\alpha,1,\cdots,P_{j-1},\cdots,P_{j},\cdots,P_{\tilde{1}},\cdots,n\big)\nonumber.
\end{align}

Notice that the sum $\gamma_{j-1+a}^\alpha(j-1)^{+a}_A+\gamma_{j+a}^\alpha(j)^{+a}_A+\gamma_{j+1+a}^\alpha(j+1)^{+a}_A+\gamma_{\tilde{1}+a}^\alpha(\tilde{1})^{+a}_A$ is invariant under the shift \eqref{shifts}. Note also that the first three delta functions become invariant under the shift after the Grassmann integrations.

This first of all means that for $d=2$ the shift is in fact a symmetry of the expression as expected because $\widetilde{\mathcal{F}}^{MHV}_{\mathcal{T}_2}$ do not depend on any Grassmann variables. For $d>2$ there will some additional Grassmann variables in $\widetilde{\mathcal{F}}^{MHV}_{\mathcal{T}_d}$. We can use conservation of super-momentum to write this function without any explicit dependence on either $\eta_{P_{j-1}}$ or $\eta_{P_{j}}$. Subsequently, we find the term proportional to $\eta_{P_{\tilde{1}}-a}\epsilon^{ab}\eta_{P_{\tilde{1}}-b}$ as well as similar factors for all other directions that has been made light-like as part of the cut. From a Feynman diagram perspective we know that such a term should always be present. The integration over the variables $\gamma_a$ can be used to remove this factor:

\begin{align}
\int d^4\eta_{P_{\tilde{1}}}\int d^2\gamma\delta^8\left(\gamma_{\tilde{1}+a}^\alpha(\tilde{1})^{+a}_A-\sum_r\eta_{rA}\lambda_r^\alpha-\eta_{P_{\tilde{1}}A}\lambda_{P_{\tilde{1}}}^\alpha\right)\eta_{P_{\tilde{1}}-a}&\epsilon^{ab}\eta_{P_{\tilde{1}}-b}\label{fermi3}\\
=2{\bf (j\tilde{1})}\langle P_jP_{j-1}\rangle^2&\delta^4\left(\langle P_{\tilde{1}}\hat{\gamma}_{\tilde{1}+a}\rangle-\sum_r\eta_{rA}\langle P_{\tilde{1}}r\rangle\right).\nonumber
\end{align}

By imposing shift symmetries for all point made light-like separated from $x_j$ (apart from $x_{j-1}$ and $x_{j+1}$), $\widetilde{\mathcal{F}}^{MHV}_{\mathcal{T}_d}$ is reduced to purely bosonic factors. This can therefore be written as $\widetilde{\mathcal{F}}^{MHV}_{\mathcal{T}_2}$ multiplied by some spinor products.

Potentially, it should be possible to use generalized unitarity to find the correct spinor factor in a systematic way by exploiting relations like the ones found in \cite{Penante:2014sza}. However, we will instead use that we already found this factor for the scalar polygon in section \ref{afsnit skalar}. Combining the information gained from the two approaches, we get:

\begin{align}
\lim_{(x_i-y_{\tilde{1}})^2=0}\frac{(x_i-y_{\tilde{1}})^2}{\bf (i\tilde{1})}&\int d^4\theta_{\tilde{1}}\begin{minipage}{2.5cm}
\includegraphics[scale=0.8]{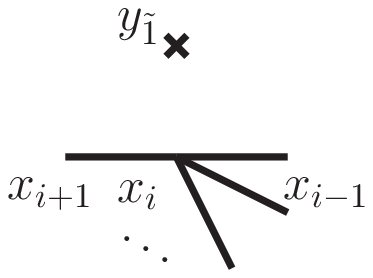}
\end{minipage}\label{vertex relation susy}\\
&=\frac{\langle i-1i\rangle}{\langle i-1\tilde{1}\rangle\langle \tilde{1}i\rangle}\frac{1}{{\bf (i\tilde{1})}}\int d^4\theta_{\tilde{1}}\delta^2(\langle \tilde{1}\theta_{\tilde{1}}^{+a}\rangle-\langle \tilde{1}\theta_i^A\rangle(\tilde{1})_A^{+a})\begin{minipage}{2.5cm}
\includegraphics[scale=0.8]{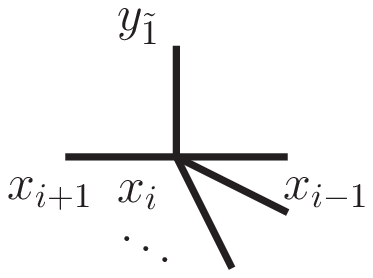}\nonumber.
\end{minipage}
\end{align}

Again full lines represent distances made light-like after dividing out scalar propagators, and vertices where $d$ lines meet correspond to an operator $\mathcal{T}_d$. As we used that all of the Grassmann variables from $\widetilde{\mathcal{F}}_{\mathcal{T}_d}^{MHV}$ in \eqref{fermi2} were removed by imposing shift symmetries, it is assumed that there are delta functions similar to \eqref{deltafunktion} for all but two of the lines meeting at $x_i$.

We will now apply all this to a full cut. A string of $m$ Lagrangian insertions will be made light-like separated from each other and the polygon such that the inside of the polygon is split into two with $x_i$ being light-like separated from $y_{\tilde{1}}$ and $x_j$ from $y_{\tilde{m}}$. In terms of the diagram in figure \ref{cut full susy}(a), the cut can be defined as:

\begin{figure}
\begin{center}
\begin{tabular}{cc}
\includegraphics[scale=1]{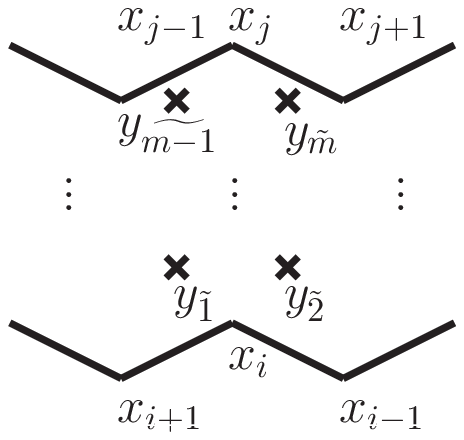}&\includegraphics[scale=1]{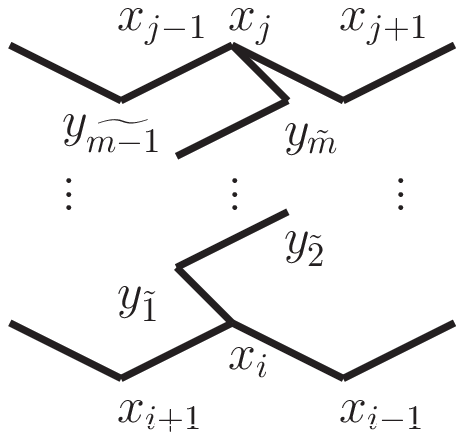}\\
(a)&(b)
\end{tabular}
\end{center}
\caption{Full lines represent distances made light-like after dividing by scalar propagators, vertices where $d$ lines meet represent operators of the type $\mathcal{T}_d$\label{cut full susy}. The polygon extends further to the left and the right in both diagrams}
\end{figure}

\begin{align}
\mathbf{Cut}=&{}\lim_{(x_i-y_{\tilde{1}})^2=0}\frac{(x_i-y_{\tilde{1}})^2}{\bf (i\tilde{1})}\lim_{(y_{\tilde{1}}-y_{\tilde{2}})^2=0}\cdots\lim_{(y_{\tilde{m}}-x_j)^2=0}\frac{(y_{\tilde{m}}-x_j)^2}{\bf (\tilde{m}j)}\int\prod_{r=1}^md^4\theta_{y_{\tilde{r}}}{\bf Fig\ref{cut full susy}(a)}.
\end{align}

We define spinors $\lambda_{\tilde{r}}^\alpha$ with $r$ going from 1 to $m+1$ ordered such that $\lambda_{\tilde{1}}^\alpha$ is a spinor corresponding to the light-like distance $x_i-y_{\tilde{1}}$ and $\lambda_{\widetilde{m+1}}^\alpha$ is a spinor corresponding to the distance $y_{\tilde{m}}-x_j$ and introduce the factor:

\begin{align}
1=&{}\frac{1}{{\bf(i\tilde{1})}}\left(\prod_{r=1}^{m-1}\frac{1}{{\bf(\tilde{r}\widetilde{r+1})}}\right)\frac{1}{{\bf(\tilde{m}j)}}\int \prod_{r=1}^{m+1}d^4\chi_r\delta^2(\chi^a_{\tilde{1}/i}-\chi^A_{\tilde{1}}(i)^{+a}_A)\delta^2(\chi^a_{\tilde{1}/\tilde{1}}-\chi^A_{\tilde{1}}(\tilde{1})^{+a}_A)\\
&\delta^2(\chi^a_{\tilde{2}/\tilde{1}}-\chi^A_{\tilde{2}}(\tilde{1})^{+a}_A)\cdots \delta^2(\chi^a_{\widetilde{m+1}/j}-\chi_{\widetilde{m+1}}^A(j)^{+a}_A).\nonumber
\end{align}

Finally we use \eqref{vertex relation susy} to write the cut in terms of the diagram in figure \ref{cut full susy}(b) and phrase it in variables common to scattering amplitudes using some of the identities found in appendix \ref{useful}:

\begin{align}
\mathbf{Cut}=&{}\frac{1}{{\bf(i\tilde{1})}}\left(\prod_{r=1}^{m-1}\frac{1}{{\bf(\tilde{r}\widetilde{r+1})}}\right)\frac{1}{{\bf(\tilde{m}j)}}\frac{\langle i-1i\rangle}{\langle i-1\tilde{1}\rangle\langle \tilde{1}\tilde{2}\rangle\cdots\langle \widetilde{m+1}j\rangle}\frac{\langle j-1j\rangle}{\langle j-1\tilde{1}\rangle\langle \tilde{1}\tilde{2}\rangle\cdots\langle \widetilde{m+1}i\rangle}\label{Cut final}\\
&\int\left(\prod_{r=1}^{m+1}d^4\eta_{\tilde{r}}\right)\delta^8\left(\sum_{r=1}^{m+1}\eta_{\tilde{r}}\lambda_{\tilde{r}}+\sum_{s=j}^{i-1}\eta_s\lambda_s\right){\bf Fig\ref{cut full susy}(b)}.\nonumber
\end{align}

\begin{figure}
\begin{center}
\includegraphics[scale=1]{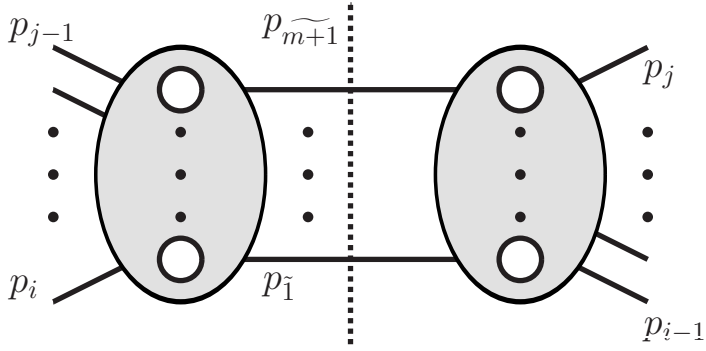}
\end{center}
\caption{The generalized unitarity cut corresponding to the cut in figure \ref{cut full susy}\label{GU cut}}
\end{figure}

When reconstructing the part of the correlation function with the propagators corresponding to the cut, the products of the harmonic variables are removed (from a generalized unitarity perspective they correspond to normalizations of the external states). The rest of \eqref{Cut final} is exactly equivalent to a generalized unitarity cut with $m+1$ cut propagators as shown in figure \ref{GU cut}. The spinor products appearing in \eqref{Cut final} correspond to the generalized unitarity cut of an MHV amplitude while everything beyond MHV lies in figure \ref{cut full susy}(b).

Equation \eqref{vertex relation susy} do not depend on the number of light-like lines meeting at the point $x_i$. Together with planarity, this allows us to do the same steps as above for each of the individual patches separated by the string of light-like propagator. This will correspond to cutting the amplitudes on either side of the generalized unitarity cut in figure \ref{GU cut}. The calculation is not going to be different from the one above, and it is straightforward to show that it will correspond to the correct generalized unitarity cut.

For the sake of completeness, let us point out that nowhere in the calculation leading up to the supersymmetric generalization in \eqref{vertex relation susy} did we use that the operator at $y_{\tilde{1}}$ was the highest fermionic component of the multiplet. The calculation can therefore be generalized to limits where the distance $x_i-x_j$ becomes light-like. In such cases the relevant delta functions will be:

\begin{align}
\delta^2(\chi_{\tilde{1}/j}-\langle\tilde{1}\theta_i^A\rangle(j)^{+a}_A)\delta^2(\chi_{\tilde{1}/i}-\langle\tilde{1}\theta^A_j\rangle(i)^{+a}_A),
\end{align}
\noindent where some of the Grassmann variables should be interpreted in terms of the specific products that appear in the construction of the correlation function, just as in \eqref{interpret}.

\subsection{Cut-Constructibility}\label{cc afsnit}

In section \ref{LI afsnit} we used operator product expansion arguments from \cite{Eden:2011yp,Eden:2011we} to conclude that the correlation functions were cut-constructible. Here we will examine one of the arguments using generalized unitarity cuts and consider the possibility of position space cuts that have no equivalent generalized unitarity. As in section \ref{susy afsnit}, it is sufficient to consider generalized unitarity cuts where light-like distance have a corresponding cut propagator.


\begin{figure}
\begin{center}
\begin{tabular}{cc}
\includegraphics[scale=0.8]{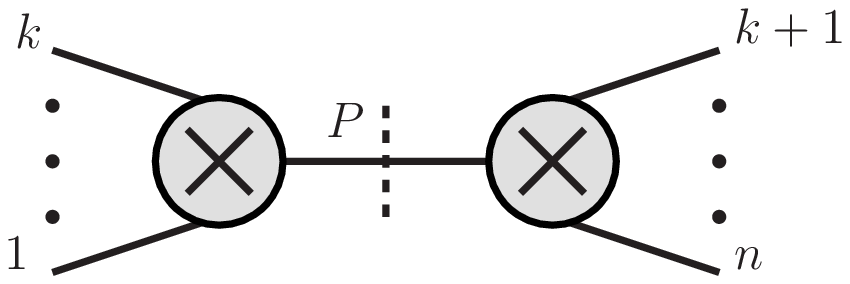}&\includegraphics[scale=0.8]{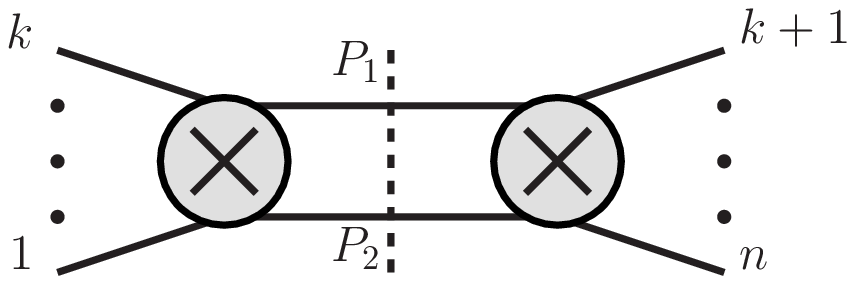}\\
(a)&(b)
\end{tabular}
\end{center}
\caption{Form factors connected through either one or two cut propagators\label{connect}}
\end{figure}

We begin by studying the order of the poles. The first case to consider is two form factors connected through a single cut propagator as shown in figure \ref{connect} (a):

\begin{align}
\int d^4\eta_P&\frac{\delta^8\left((1)^{+a}_{A}\gamma_{1+a}^\alpha-\sum_{r=1}^k\eta_{rA}\lambda_r^\alpha-\eta_{PA}\lambda_P^\alpha\right)}{\prod_{i=1}^{k-1}\langle ii+1\rangle \langle kP\rangle\langle P1\rangle}\frac{\delta^8\left((2)^{+a}_{A}\gamma_{2+a}^\alpha-\sum_{s=k+1}^n\eta_{sA}\lambda_s^\alpha+\eta_{PA}\lambda_P^\alpha\right)}{\prod_{j=k}^{n-1}\langle jj+1\rangle\langle nP\rangle\langle Pk\rangle}\\
&=\frac{\delta^8\left((1)^{+a}_{A}\gamma_{1+a}^\alpha+(2)^{+a}_{A}\gamma_{2+a}^\alpha-\sum_{r=1}^n\eta_{rA}\lambda_r^\alpha\right)}{\prod_{i=1}^{k-1}\langle ii+1\rangle\prod_{j=k+1}^{n-1}\langle jj+1\rangle}\frac{\delta^4\left((2)^{+a}_{A}\langle\gamma_{2+a}P\rangle-\sum_{s=P+1}^n\eta_{sA}\langle sP\rangle\right)}{\langle kP\rangle\langle Pk+1\rangle\langle P1\rangle\langle nP\rangle}\nonumber.
\end{align}

From this we see that there are as many factors of $P$ in the numerator as in the denominator. Including the cut propagator, there are then one more propagator depending on $P$ than momentum factors in the numerator. This is exactly enough to give the divergence of a scalar propagator ($i.e.$ a simple pole in $(x_1-x_2)^2$) provided all the momentum factors become derivatives with respect to $x_1$ or $x_2$. These derivatives will then give something proportional to $(x_1-x_2)^\mu$ (see section \ref{div Wilson}).

The second case to consider is two form factors connected through two cut propagators as shown in figure \ref{connect} (b):

\begin{align}
&\int d^4\eta_{P_1}d^4\eta_{P_2}\frac{\delta^8\left((1)^{+a}_{A}\gamma_{1+a}^\alpha-\sum_{r=1}^k\eta_{rA}\lambda_r^\alpha-\eta_{P_1A}\lambda_{P_1}^\alpha-\eta_{P_2A}\lambda_{P_2}^\alpha\right)}{\prod_{i=1}^{k-1}\langle ii+1\rangle \langle kP_1\rangle\langle P_1P_2\rangle\langle P_21\rangle}\nonumber\\
&\frac{\delta^8\left((2)^{+a}_{A}\gamma_{2+a}^\alpha-\sum_{s=k+1}^n\eta_{sA}\lambda_s^\alpha+\eta_{P_1A}\lambda_{P_1}^\alpha+\eta_{P_2A}\lambda_{P_2}^\alpha\right)}{\prod_{j=k}^{n-1}\langle jj+1\rangle\langle nP_2\rangle\langle P_2P_1\rangle\langle P_1k+1\rangle}\\
&\,\,\,\,\,\,\,\,\,\,\,\,\,\,\,\,\,\,\,\,=-\frac{\delta^8\left((1)^{+a}_{A}\gamma_{1+a}^\alpha+(2)^{+a}_{A}\gamma_{2+a}^\alpha-\sum_{r=1}^n\eta_{rA}\lambda_r^\alpha\right)}{\prod_{i=1}^{k-1}\langle ii+1\rangle\prod_{j=k+1}^{n-1}\langle jj+1\rangle}\frac{\langle P_1P_2\rangle^2}{\langle kP_1\rangle\langle P_1k+1\rangle\langle nP_2\rangle\langle P_21\rangle}\nonumber.
\end{align}

Both cut propagators could give simple poles in position space. But because the $P_1$ spinors are contracted with the $P_2$ spinors, it is not possilbe to a double pole. Attempting to create a double pole will give a factor:

\begin{align*}
(\cancel{x}_1-\cancel{x}_2)(\cancel{x}_1-\cancel{x}_2),
\end{align*}
\noindent in the numerator which would lower the divergence to a simple pole. The exception is when there are no on-shell states apart from $P_1$ and $P_2$. This correspond to disconnected graphs. We thus arrive at the same conclusion as the operator product expansion gave us. The correlation functions contain simple poles and double poles, and the double poles correspond to disconnected diagrams.

\begin{figure}
\begin{center}
\begin{tabular}{cc}
\includegraphics[scale=0.7]{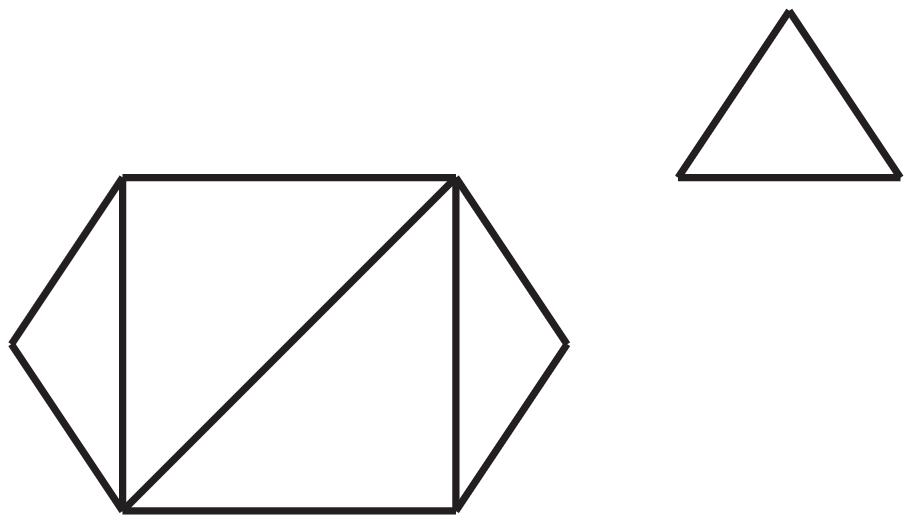}&\includegraphics[scale=0.7]{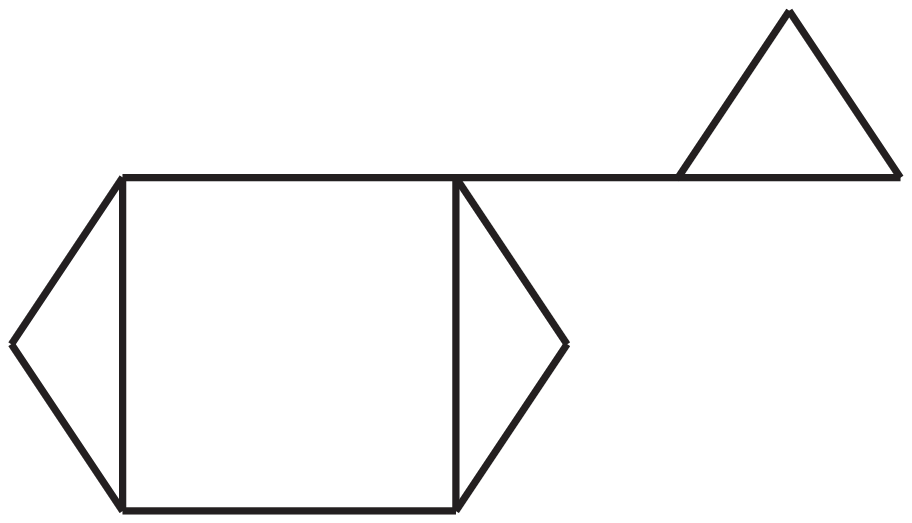}
\end{tabular}
\end{center}
\caption{Position space cuts with no equivalent amplitudes generalized unitarity cuts. Full lines represent distances made light-like after dividing by the appropriate propagators\label{problematisk figur}}
\end{figure}

Let us now move on to the position space cuts that do not have equivalent generalized unitarity cuts. This could be terms where a group of Lagrangian insertion points only connect among themselves or only once to a point on the polygon as shown in figure \ref{problematisk figur}. We expect such terms to correspond to unconnected diagrams and diagrams proportional to a group structure constant with two identical indices. Below we will argue that this expectation holds.

For the relevant generalized unitarity cuts every light-like distance have a corresponding cut propagator. The converse can also be made true in a sense. As shown above two form factors connected through a cut propagator contain exactly the right number of momentum factors to give a simple pole. Of course as shown in section \ref{afsnit skalar}, this does not guarantee that the pole appears. Still this relies on several non-light-like distances ruining the divergences for each other, and in a planar diagram at least one will create a simple pole.

Let us then consider one of the unwanted terms with a certain set of simple poles and find the residue for all the possible limits:
\begin{align*}
(x_i-x_j)^2&=0,&(x_i-y_{\tilde{\jmath}})^2&=0,&(y_{\tilde{\imath}}-y_{\tilde{\jmath}})^2&=0.
\end{align*}

Then the cut propagators in the relevant generalized unitarity cuts will either all have a corresponding light-like distance or there will be an additional simple pole. The latter contradict the assumption that we took all the possible limits.

All cut propagators in the relevant generalized unitarity cuts then correspond to light-like distances. Hence the only generalized unitarity cuts, we can write down for the unwanted position space cuts, match the expectation and are not relevant to the duality.

\section{More General Correlators}\label{General afsnit}

Let us finally turn towards other correlators as well as non-planar contributions and discuss how they can be computed.

One type to consider is correlation functions with both operators from the stress-tensor multiplet arranged in a light-like polygon and other operators not part of the polygon. These correlators are a natural extension to the duality between correlation functions and Wilson loops as the light-like limit simply gives the correlation function of a Wilson loop and the additional operators \cite{Alday:2011ga}. The additional operators can also be arranged to form a second Wilson loop.

\begin{figure}
\begin{center}
\includegraphics[scale=1]{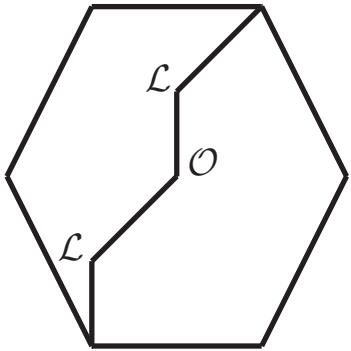}
\end{center}
\caption{Full lines represent distances made light-like after dividing by the appropriate propagators. $\mathcal{O}$ represents some arbitrary operator\label{extra figur}}
\end{figure}

It is still possible to define position space cuts for such correlators even though there is no duality with amplitudes. These cuts will include diagrams where Lagrangian insertion points are made light-like separated from the additional operators as shown in figure \ref{extra figur}. It is not clear if such correlators will be cut-constructible, something that may well depend on the specific choice of operators. Adding a single operator to a light-like polygon could be a good starting point for considering correlation functions of other operators as many details will be similar to the light-like polygon. In the case of the additional operators forming a second Wilson loop the cuts will be similar to the ones used for the duality and can be computed from \eqref{vertex relation susy}.

\begin{figure}
\begin{center}
\includegraphics[scale=0.5]{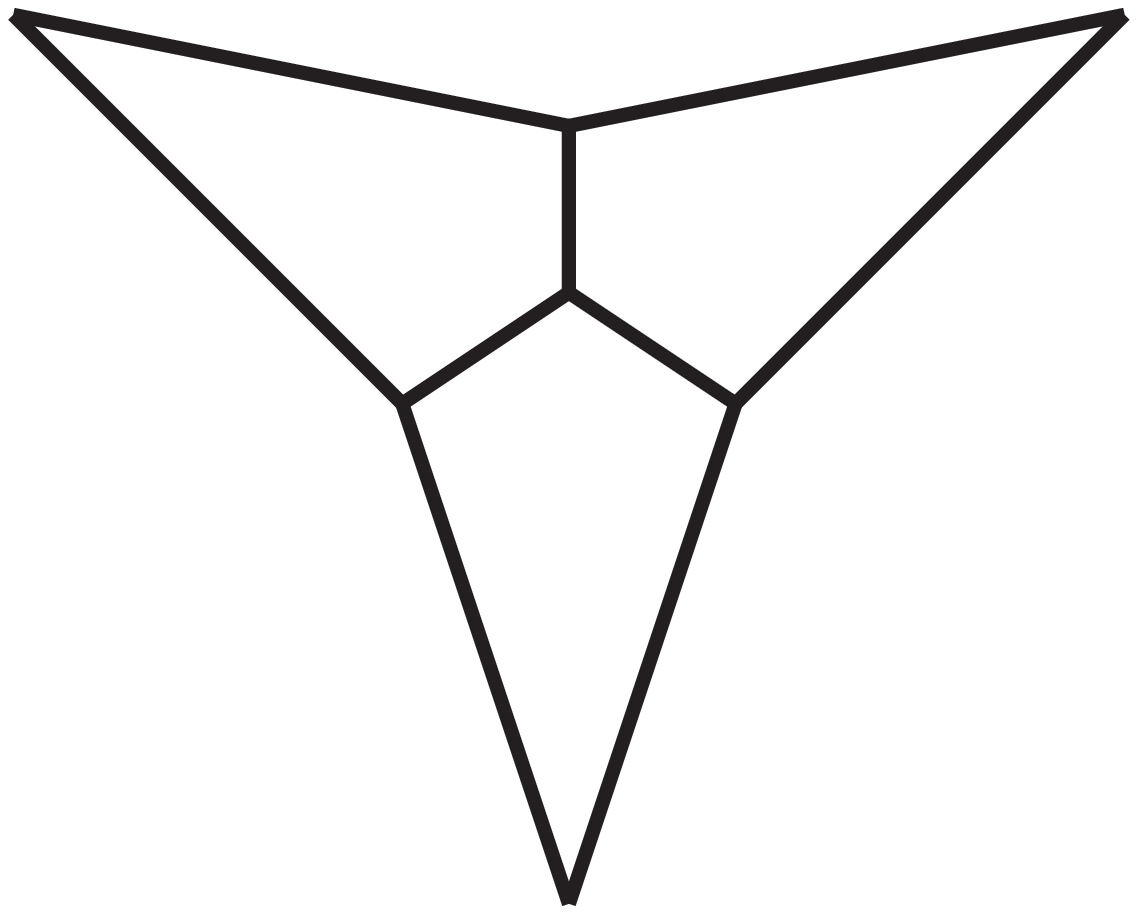}
\end{center}
\caption{Full lines represent distances made light-like after dividing by scalar propagators, vertices where $d$ lines meet represent operators of the type $\mathcal{T}_d$\label{ny cf}}
\end{figure}

One way to construct other correlators that are cut-constructible and also dual to scattering amplitudes is to consider operators of the type half-BPS operators as in \eqref{def T_k} with $d>2$ as shown in figure \ref{ny cf}. This sort of diagram will appear as part of the cuts used in section \ref{Duality afsnit} but one could also use this as the starting point. Since equations \eqref{fermi1} and \eqref{fermi2} do not rely on integration over the super-space variables, it should be dual to three different four-point amplitudes and a single six-point amplitude provided we introduce some additional fermionic delta functions like the ones in \eqref{deltafunktion}.

\subsection{Operators at Generic Points}\label{generic points afsnit}

For the duality, the original correlation functions already involved some light-like limit. It would be interesting to compute the cuts for the correlator with operators at generic points. The arguments reviewed in section \ref{LI afsnit} are sufficient to argue that the correlation function is cut-constructible. However, the position space cuts may not be as easy to compute as for those correlators appearing in the duality.

\begin{figure}
\begin{center}
\includegraphics[scale=1]{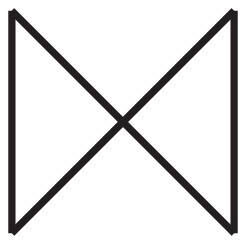}
\end{center}
\caption{Full lines represent distances made light-like after dividing by scalar propagators, vertices where $d$ lines meet represent operators of the type $\mathcal{T}_d$\label{extra cut}}
\end{figure}

For the correlation function of only four purely scalar operators the integrand is known to a high loop order \cite{Eden:1998hh,GonzalezRey:1998tk,Eden:2000mv,Bianchi:2000hn,Eden:2011we}. We can use the results to check whether the correlators can be constructed from cuts and what those cuts are. The one-loop integrand consists of two types of terms: those that contribute to the duality between scattering amplitudes and correlation functions and terms that can be captured by cuts where all distances to the Lagrangian insertion point become light-like while the original operators only become light-like separated from each other in pairs. The new cuts are proportional to the correlation function where the Lagrangian insertion has been replaced by the lowest component of $\mathcal{T}_4$ as in figure \ref{extra cut}.

\begin{figure}
\begin{center}
\begin{tabular}{ccc}
\includegraphics[scale=1]{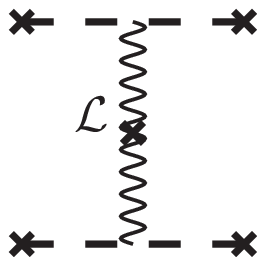}&\includegraphics[scale=1]{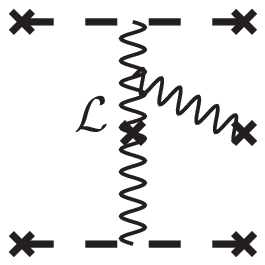}&\includegraphics[scale=1]{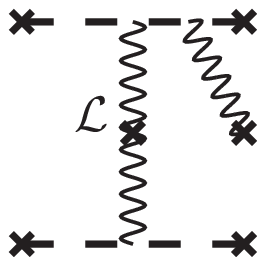}\\
(a)&(b)&(c)
\end{tabular}
\end{center}
\caption{Feynman diagrams with a Lagrangian insertion in the center interacting with operators with scalars. We do not show the entire Feynman diagrams only the relevant parts\label{4-pt cut}}
\end{figure}

If this is to make sense as a cut, we require that for similar limits at higher loop orders, the Lagrangian insertion should still act like four scalars. This is obviously true for the part of the on-shell Lagrangian proportional to four scalars but the arguments from section \ref{afsnit skalar} can be used to argue that it will also be the case for the other parts. Consider for instance the diagrams in figure \ref{4-pt cut} and the limit where the Lagrangian in the center of the diagrams is made light-like separated from the four operators at the corners of the diagrams. The diagram in \ref{4-pt cut}(a) contribute to the mentioned limit, but the diagram in \ref{4-pt cut}(b) do not without some additional light-like limit involving two of the original five operators, whereas \ref{4-pt cut}(c) do contribute without any additional limits. This systematic continues with more interactions. Only as long as the additional interactions are with the scalar lines will the diagrams contribute to the aforementioned limit without the need for any additional light-like limits involving two or more of the original five operators. For this reason we may conclude that in this limit the Lagrangian insertion acts like four scalars. Consequently, the cut can be described in terms of the correlation function where the Lagrangian insertion has been replaced by the lowest fermionic component of $\mathcal{T}_4$.

By inspecting the results from the literature, we see that at higher loop orders, it is always possible to make the insertion points light-like separated from four other points and let the points of the original operators be light-like separated from each other in pairs. So up to the known loop order, the correlation function should be determined by the cuts if we include cuts where the Lagrangian insertions get replaced by $\mathcal{T}_4$. These new cuts may appear identical to applying \eqref{vertex relation susy} twice. Indeed if a Lagrangian insertion is made light-like separated from two purely scalar operators using this relation, the result would be proportional to a correlation function with $\mathcal{T}_4$ in place of the on-shell Lagrangian. The difference lies in the fact that for these new cuts the operators made light-like from the Lagrangian insertion may be light-like separated from only one other operator. As a consequence, the factor being pulled out in front of the correlation function, when doing the cut, is different. For the relation \eqref{vertex relation susy} this factor consists of spinor products and could be found through a fairly simple Feynman diagram calculation. The computation needed for the additional cuts do not seem to be as simple, and the factor would involve tensors connecting the $\mathrm{SU(4)}$ indices of the harmonic variables.

\subsection{Non-Planar Diagrams}

The duality discussed in setion \ref{Duality afsnit} considers only planar diagrams so we only formulated cuts for the planar theory. However, it is also possible to consider non-planar cuts. Equations \eqref{fermi1} and \eqref{fermi2} do in fact not rely on planarity so the Grassmann structure of \eqref{vertex relation susy} will be the same in the non-planar case. The kinematical factor can again be found from considering purely scalar operators, and the calculations will be very similar to the one leading to \eqref{vertex relation}. For the sake of clarity, we only display the result with a limited number of operators though the generalization is straightforward. We have introduced an additional point $x_j$ and defined spinors such that $(x_i-x_j)^{\alpha\dot{\alpha}}=\lambda_{j}^\alpha\tilde{\lambda}^{\dot{\alpha}}_{j}$, the limit needed for the cuts is then given by:

\begin{align}
\lim_{(x_i-y_{\tilde{1}})^2=0}\frac{(x_i-y_{\tilde{1}})^2}{\bf (i\tilde{1})}&\int d^4\theta_{\tilde{1}}\begin{minipage}{2.5cm}
\includegraphics[scale=0.8]{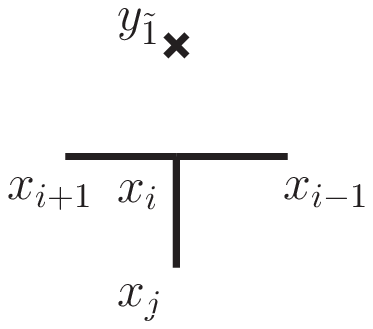}
\end{minipage}\nonumber\\
=&{}\frac{1}{{\bf (i\tilde{1})}}\int d^4\theta_{\tilde{1}}\delta^2(\langle \tilde{1}\theta_{\tilde{1}}^{+a}\rangle-\langle \tilde{1}\theta_i^A\rangle(\tilde{1})_A^{+a})\Bigg[\frac{\langle i-1i\rangle}{\langle i-1\tilde{1}\rangle\langle \tilde{1}i\rangle}\begin{minipage}{2.5cm}
\includegraphics[scale=0.8]{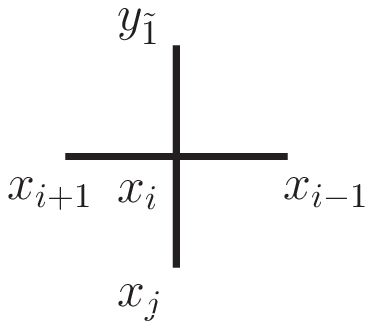}
\end{minipage}\\
&+\frac{\langle ji-1\rangle}{\langle j\tilde{1}\rangle\langle \tilde{1}i-1\rangle}\begin{minipage}{3cm}
\includegraphics[scale=0.8]{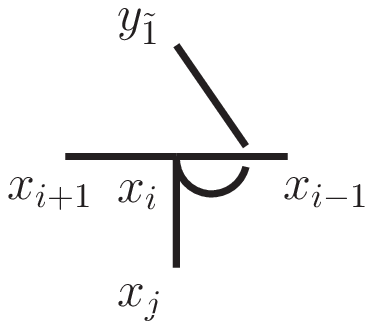}
\end{minipage}+\frac{\langle ij\rangle}{\langle i\tilde{1}\rangle\langle \tilde{1}j\rangle}\begin{minipage}{3cm}
\includegraphics[scale=0.8]{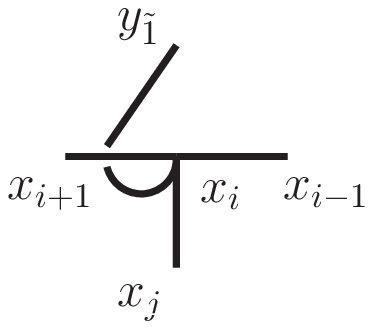}
\end{minipage}\Bigg].\nonumber
\end{align}

One should note that outside the planar limit the cuts no longer separate the diagrams into separate patches as in section \ref{Duality afsnit}, though the cuts may still simplify the expression.

\section{Discussion}\label{Diskussionsafsnit}

In these notes, we have introduced a notion of cuts in position space and shown how this type of cuts on a specific set of limits of correlation functions correspond to generalized unitarity cuts of scattering amplitudes. This means that the super-correlators/super-amplitudes duality works on a cut-by-cut basis. We also checked that the super-correlators, considered in the duality, are in fact completely determined by the position space cuts. The results are hardly surprising as the supersymmetric correlation functions have been found to be dual to the supersymmetric Wilson loop of \cite{Mason:2010yk}, and since the duality between scattering amplitudes and correlation functions is at the integrand level, no regularization issues should arise. Nonetheless, it provided a simple example to try out this reformulation of the Lagrangian insertion technique.

The position space cuts are written in terms of correlation functions of other half-BPS operators but there is a non-trivial factor that emerges from doing the cut unlike for generalized unitarity where all the non-trivial information lies in the product of amplitudes. This might be a problem for more general correlation functions like the ones considered in section \ref{generic points afsnit}. We identified a second type of cuts still written in terms of correlators of half-BPS operators but with factors that do not follow as easily as for the cuts used in the duality.

In general, it would be interesting to extend this approach to other operators. It will certainly be possible to define the cuts but it is not clear if the correlators will be cut-constructible nor whether the cuts will be simple. The extension to non-planar diagrams is more straightforward though the cuts will no longer divide the Feynman diagrams into separate patches.

Since the generalized unitarity methods are related the twistor space methods for scattering amplitudes, we expect that this approach is related to the twistor space methods used in \cite{Adamo:2011dq,Chicherin:2014uca}, and it would be interesting to find the direct relation.

\section*{Acknowledgments}

I have benefited from discussions with Henrik Johansson and Radu Roiban and am grateful for useful comments on an early draft by Henrik Johansson and Gregory Korchemsky. This work is supported by the Knut and Alice Wallenberg Foundation under grant KAW~2013.0235. Figures in these notes were drawn using JaxoDraw \cite{Binosi:2003yf}.

\appendix

\section{Harmonic Variables and Spinors}\label{notationsappendix}

In this appendix we briefly sum up some of the conventions and notations used. The harmonic variables are matrices satisfying the following relations:

\begin{align}
\bar{u}_{+a}^{A}u_{A}^{+b}=\delta^b_a,\quad\bar{u}_{-a'}^{A}u_{A}^{-b'}&=\delta^{b'}_{a'},
&\bar{u}_{-a'}^{A}u_{A}^{+b}=\bar{u}_{+a}^{A}u^{-b'}_{A}&=0,\\
u^{+a}_{A}\bar{u}_{+a}^{B}+u^{-a'}_{A}\bar{u}_{-a'}^{B}&=\delta^B_A,&\tfrac{1}{4}\epsilon^{ABCD}u^{+a}_{A}\epsilon_{ab}u^{+b}_{B}u^{-c'}_{C}\epsilon_{c'd'}u^{-d'}_{D}&=1.
\end{align}

Upper-case Latin indices are $\mathrm{SU(4)}$ indices while lower-case Latin indices are $\mathrm{SU(2)}$ indices. Since we will be dealing with operators at many different points it is convenient to use a notation that makes for an easy identification of the corresponding harmonic variables for each operator: we choose to denote the harmonic variables of the operator at point $x_i$ on the polygon by $(i)^{+a}_{A}$ and the harmonic variables of the Lagrangian insertion at point $y_{\tilde{m}}$ by $(\tilde{m})^{+a}_{A}$. It is also useful to introduce this product of harmonic variables:

\begin{align}
\mathbf{(ij)}&=\tfrac{1}{4}\epsilon^{ABCD}(i)^{+a}_{A}\epsilon_{ab}(i)^{+b}_{B}(j)^{+c}_{C}\epsilon_{cd}(j)_{D}^{+d}.\label{harmony}
\end{align}

The product correspond to the determinant of the matrix:

\begin{align}
(\overline{\imath}j)_{a'}^{\phantom{a'}a}=(\overline{\imath})_{-a'}^A(j)_A^{+a}.
\end{align}

As long as the product \eqref{harmony} is non-zero it is possible to define the inverse matrix $(\overline{\imath}j)_{\phantom{-1}a}^{-1\phantom{a}a}$ and by expressing $\mathrm{SU(4)}$ vectors in terms of $(i)^{+a}_{A}$ and $(j)^{+a}_{A}$ it is possible to show that the following is the identity matrix:

\begin{align}
(j)^{+a}_B(\overline{\imath}j)_{\phantom{-1}a}^{-1\phantom{a}a'}(\overline{\imath})_{-a'}^A
+(i)^{+a}_B(\overline{\jmath}i)_{\phantom{-1}a}^{-1\phantom{a}a'}(\overline{\jmath})_{-a'}^A&=\delta_B^A.
\end{align}

We use greek letters from the beginning of the alphabet for spinor indices while $\mu$ and $\nu$ are reserved for regular Lorentz indices. The spinor indices are raised and lowered as follows:

\begin{align}
\lambda^\alpha&=\epsilon^{\alpha\beta}\lambda_\beta,&\lambda_\alpha&=\epsilon_{\alpha\beta}\lambda^\beta,
\end{align}
\noindent while the spinor product is defined to be:

\begin{align}
\langle ij\rangle&=\lambda_i^\alpha\lambda_{j\alpha}.
\end{align}

The Levi-Civita symbols are chosen to be:

\begin{align}
\epsilon_{12}=\epsilon_{\dot{1}\dot{2}}=&1=\epsilon^{21}=\epsilon^{\dot{2}\dot{1}}.
\end{align}

Lorentz vectors can be written in spinor notation by using Pauli matrices:

\begin{align}
x_{\alpha\dot{\alpha}}=&{}\sigma^\mu_{\alpha\dot{\alpha}}x_\mu.
\end{align}

\section{Integrals}\label{integral app}

The following quantity is useful when describing the propagation from a point on a light-like Wilson line to some point $y_{\tilde{\jmath}}$:

\begin{align}
\Delta(y_{\tilde{\jmath}},t_j)=&{}\frac{1}{(x_{i+1}-y_{\tilde{\jmath}})^2(1-t_j)+(x_i-y_{\tilde{\jmath}})^2t_j}.
\end{align}

For convenience we also define the following quantities:

\begin{align}
\Upsilon_{\mu_j\nu_j}(y_{\tilde{\jmath}})=&{}2(x_{i+1}-y_{\tilde{\jmath}})_{[\mu_j}(x_i-x_{i+1})_{\nu_j]}
\end{align}

\begin{align}
\Xi(y_{\tilde{\imath}},y_{\tilde{\jmath}})=&{}\frac{1}{(x_i-y_{\tilde{\imath}})^2(x_{i+1}-y_{\tilde{\jmath}})^2-(x_i-y_{\tilde{\jmath}})^2(x_{i+1}-y_{\tilde{\imath}})^2}
\end{align}

The second quantity satisfy these relations

\begin{align}
\Xi(y_{\tilde{\imath}},y_{\tilde{\jmath}})=&{}-\Xi(y_{\tilde{\jmath}},y_{\tilde{\imath}}),\\
\lim_{(x_i-y_{\tilde{\imath}})^2=0}\Xi(y_{\tilde{\imath}},y_{\tilde{\jmath}})=&{}-\Delta(y_{\tilde{\jmath}},1)\Delta(y_{\tilde{\imath}},0).
\end{align}

\begin{align}
I_1(y_{\tilde{1}};t_2)=&{}\int_{t_2}^1dt_1(x_i-x_{i+1})_{[\mu_1}\left(\frac{\partial}{\partial y_{\tilde{1}}^{\nu_1]}}\Delta(y_{\tilde{1}},t_1)\right)\\
=&{}\Upsilon_{\mu_1\nu_1}(y_{\tilde{1}})(1-t_2)\Delta(y_{\tilde{1}},1)\Delta(y_{\tilde{1}},t_2)\nonumber.
\end{align}

\begin{align}
I_2(y_{\tilde{2}},y_{\tilde{1}};t_3)=&{}\int_{t_3}^1dt_2(x_i-x_{i+1})_{[\mu_2}\left(\frac{\partial}{\partial y_{\tilde{2}}^{\nu_2]}}\Delta(y_{\tilde{2}},t_2)\right)I_1(t_2)\nonumber\\
=&{}-\Upsilon_{\mu_1\nu_1}(y_{\tilde{1}})\Xi(y_{\tilde{1}},y_{\tilde{2}})\frac{\Delta(y_{\tilde{1}},1)}{\Delta(y_{\tilde{2}},1)}I_1(y_{\tilde{2}};t_3)-\Upsilon_{\mu_2\nu_2}(y_{\tilde{2}})\Upsilon_{\mu_1\nu_1}(y_{\tilde{1}})\Xi(y_{\tilde{1}},y_{\tilde{2}})^2\\
&\times\Bigg[\mathrm{ln}\left(\frac{\Delta(y_{\tilde{2}},t_3)}{\Delta(y_{\tilde{2}},1)}\right)-\mathrm{ln}\left(\frac{\Delta(y_{\tilde{1}},t_3)}{\Delta(y_{\tilde{1}},1)}\right)\Bigg]\nonumber.
\end{align}

\begin{align}
I_3(y_{\tilde{3}},y_{\tilde{2}},y_{\tilde{1}};t_4)=&{}\int_{t_4}^1dt_3(x_i-x_{i+1})_{[\mu_3}\left(\frac{\partial}{\partial y_{\tilde{3}}^{\nu_3]}}\Delta(y_{\tilde{3}},t_3)\right)I_2(y_{\tilde{2}},y_{\tilde{1}};t_3)\nonumber\\
=&{}-\Upsilon_{\mu_1\nu_1}(y_{\tilde{1}})\Xi(y_{\tilde{1}},y_{\tilde{2}})\frac{\Delta(y_{\tilde{1}},1)}{\Delta(y_{\tilde{2}},1)}I_2(y_{\tilde{3}},y_{\tilde{2}};t_4)-\Upsilon_{\mu_3\nu_3}(y_{\tilde{3}})\Upsilon_{\mu_2\nu_2}(y_{\tilde{2}})\Upsilon_{\mu_1\nu_1}(y_{\tilde{1}})\\
&\times\Xi^2(y_{\tilde{1}},y_{\tilde{2}})\Bigg[\frac{\Xi(y_{\tilde{1}},y_{\tilde{3}})\Xi(y_{\tilde{2}},y_{\tilde{3}})}{\Xi(y_{\tilde{1}},y_{\tilde{2}})}\mathrm{ln}\left(\frac{\Delta(y_{\tilde{3}},t_4)}{\Delta(y_{\tilde{3}},1)}\right)\nonumber\\
&+\frac{\Xi(y_{\tilde{1}},y_{\tilde{3}})\Delta(y_{\tilde{3}},t_4)}{\Delta(y_{\tilde{1}},t_4)}\mathrm{ln}\left(\frac{\Delta(y_{\tilde{1}},t_4)}{\Delta(y_{\tilde{1}},1)}\right)-\frac{\Xi(y_{\tilde{2}},y_{\tilde{3}})\Delta(y_{\tilde{3}},t_4)}{\Delta(y_{\tilde{2}},t_4)}\mathrm{ln}\left(\frac{\Delta(y_{\tilde{2}},t_4)}{\Delta(y_{\tilde{2}},1)}\right)\Bigg]\nonumber.
\end{align}

\begin{align}
&I_4(y_{\tilde{4}},y_{\tilde{3}},y_{\tilde{2}},y_{\tilde{1}};t_5)\nonumber\\
=&{}\int_{t_5}^1dt_4(x_i-x_{i+1})_{[\mu_4}\left(\frac{\partial}{\partial y_{\tilde{4}}^{\nu_4]}}\Delta(y_{\tilde{4}},t_4)\right)I_3(y_{\tilde{3}},y_{\tilde{2}},y_{\tilde{1}};t_4)\nonumber\\
=&{}-\Upsilon_{\mu_1\nu_1}(y_{\tilde{1}})\Xi(y_{\tilde{1}},y_{\tilde{2}})\frac{\Delta(y_{\tilde{1}},1)}{\Delta(y_{\tilde{2}},1)}I_3(y_{\tilde{4}},y_{\tilde{3}},y_{\tilde{2}};t_5)-\Upsilon_{\mu_4\nu_4}(y_{\tilde{4}})\Upsilon_{\mu_3\nu_3}(y_{\tilde{3}})\Upsilon_{\mu_2\nu_2}(y_{\tilde{2}})\Upsilon_{\mu_1\nu_1}(y_{\tilde{1}})\nonumber\\
&\times\Xi^2(y_{\tilde{1}},y_{\tilde{2}})\Bigg[-\frac{\Xi(y_{\tilde{1}},y_{\tilde{3}})\Xi(y_{\tilde{2}},y_{\tilde{3}})\Xi(y_{\tilde{3}},y_{\tilde{4}})\Delta(y_{\tilde{4}},t_5)}{\Xi(y_{\tilde{1}},y_{\tilde{2}})\Delta(y_{\tilde{3}},t_5)}\mathrm{ln}\left(\frac{\Delta(y_{\tilde{3}},t_5)}{\Delta(y_{\tilde{3}},1)}\right)\\
&-\frac{\Xi(y_{\tilde{1}},y_{\tilde{3}})\Xi(y_{\tilde{3}},y_{\tilde{4}})\Delta(y_{\tilde{4}},t_5)}{\Delta(y_{\tilde{1}},t_5)}\mathrm{ln}\left(\frac{\Delta(y_{\tilde{1}},t_5)}{\Delta(y_{\tilde{1}},1)}\right)+\frac{\Xi(y_{\tilde{2}},y_{\tilde{3}})\Xi(y_{\tilde{3}},y_{\tilde{4}})\Delta(y_{\tilde{4}},t_5)}{\Delta(y_{\tilde{2}},t_5)}\mathrm{ln}\left(\frac{\Delta(y_{\tilde{2}},t_5)}{\Delta(y_{\tilde{2}},1)}\right)\nonumber\\
&+\Xi^2(y_{\tilde{3}},y_{\tilde{4}})\bigg(\mathrm{Li}_2\left(\frac{(\Delta(y_{\tilde{3}},1)-\Delta(y_{\tilde{3}},0))\Xi(y_{\tilde{1}},y_{\tilde{3}})}{\Delta(y_{\tilde{3}},1)\Delta(y_{\tilde{3}},0)\Delta(y_{\tilde{1}},t_5)}\right)-\mathrm{Li}_2\left(\frac{(\Delta(y_{\tilde{3}},1)-\Delta(y_{\tilde{3}},0))\Xi(y_{\tilde{1}},y_{\tilde{3}})}{\Delta(y_{\tilde{3}},1)\Delta(y_{\tilde{3}},0)\Delta(y_{\tilde{1}},1)}\right)\nonumber\\
&-\mathrm{Li}_2\left(\frac{(\Delta(y_{\tilde{4}},1)-\Delta(y_{\tilde{4}},0))\Xi(y_{\tilde{1}},y_{\tilde{4}})}{\Delta(y_{\tilde{4}},1)\Delta(y_{\tilde{4}},0)\Delta(y_{\tilde{1}},t_5)}\right)+\mathrm{Li}_2\left(\frac{(\Delta(y_{\tilde{4}},1)-\Delta(y_{\tilde{4}},0))\Xi(y_{\tilde{1}},y_{\tilde{4}})}{\Delta(y_{\tilde{4}},1)\Delta(y_{\tilde{4}},0)\Delta(y_{\tilde{1}},1)}\right)\nonumber\\
&-\mathrm{Li}_2\left(\frac{(\Delta(y_{\tilde{3}},1)-\Delta(y_{\tilde{3}},0))\Xi(y_{\tilde{2}},y_{\tilde{3}})}{\Delta(y_{\tilde{3}},1)\Delta(y_{\tilde{3}},0)\Delta(y_{\tilde{2}},t_5)}\right)+\mathrm{Li}_2\left(\frac{(\Delta(y_{\tilde{3}},1)-\Delta(y_{\tilde{3}},0))\Xi(y_{\tilde{2}},y_{\tilde{3}})}{\Delta(y_{\tilde{3}},1)\Delta(y_{\tilde{3}},0)\Delta(y_{\tilde{2}},1)}\right)\nonumber\\
&+\mathrm{Li}_2\left(\frac{(\Delta(y_{\tilde{4}},1)-\Delta(y_{\tilde{4}},0))\Xi(y_{\tilde{2}},y_{\tilde{4}})}{\Delta(y_{\tilde{4}},1)\Delta(y_{\tilde{4}},0)\Delta(y_{\tilde{2}},t_5)}\right)-\mathrm{Li}_2\left(\frac{(\Delta(y_{\tilde{4}},1)-\Delta(y_{\tilde{4}},0))\Xi(y_{\tilde{2}},y_{\tilde{4}})}{\Delta(y_{\tilde{4}},1)\Delta(y_{\tilde{4}},0)\Delta(y_{\tilde{2}},1)}\right)\nonumber\\
&+\bigg[\mathrm{ln}\left(\frac{\Delta(y_{\tilde{1}},t_5)}{\Delta(y_{\tilde{1}},1)}\right)-\mathrm{ln}\left(\frac{\Delta(y_{\tilde{2}},t_5)}{\Delta(y_{\tilde{2}},1)}\right)\bigg]\bigg[\mathrm{ln}\left(\frac{\Delta(y_{\tilde{3}},t_5)}{\Delta(y_{\tilde{3}},1)}\right)-\mathrm{ln}\left(\frac{\Delta(y_{\tilde{4}},t_5)}{\Delta(y_{\tilde{4}},1)}\right)\bigg]\nonumber\\
&+\mathrm{ln}\left(\frac{\Delta(y_{\tilde{1}},t_5)}{\Delta(y_{\tilde{1}},1)}\right)\mathrm{ln}\left(\frac{\Delta(y_{\tilde{4}},1)\Xi(y_{\tilde{1}},y_{\tilde{3}})}{\Delta(y_{\tilde{3}},1)\Xi(y_{\tilde{1}},y_{\tilde{4}})}\right)-\mathrm{ln}\left(\frac{\Delta(y_{\tilde{2}},t_5)}{\Delta(y_{\tilde{2}},1)}\right)\mathrm{ln}\left(\frac{\Delta(y_{\tilde{4}},1)\Xi(y_{\tilde{2}},y_{\tilde{3}})}{\Delta(y_{\tilde{3}},1)\Xi(y_{\tilde{2}},y_{\tilde{4}})}\right)\Bigg]\nonumber.
\end{align}

\section{Jacobians and Useful Identities}\label{useful}

Changing from a measure for the fermionic variables $\theta^{+a}_{\tilde{r}\alpha}$ into a measure for the variables $\chi_{\tilde{r}/\tilde{r}}^a=\langle\tilde{r}\theta_{\tilde{r}}^a\rangle$ and $\chi_{\widetilde{r+1}/\tilde{r}}^a=\langle\widetilde{r+1}\theta_{\tilde{r}}^a\rangle$ is going to introduce the Jacobian:

\begin{align}
\prod_{r=1}^m\langle\tilde{r}\widetilde{r+1}\rangle^2.
\end{align}

The duality gives the scattering amplitudes in terms of the fermionic parts of the super-twistors, they can be related to the Grassmann variables, $\eta^A_i$, where $(\eta_i)^0$ indicates a positive helicity gluon of momentum $p_i$ and $(\eta_i)^4$ indicates a negative helicity gluon, in the following way:
\begin{align}
\eta^A_i=&{}\frac{\chi_{i-1}^A\langle ii+1\rangle+\chi_i^A\langle i+1i-1\rangle+\chi_{i+1}^A\langle i-1i\rangle}{\langle i-1i\rangle\langle ii+1\rangle}.
\end{align}

For the Grassmann variables of the internal states there are different possible definitions, we choose:

\begin{align}
\eta^A_{\tilde{1}}=&{}\frac{\chi_{i}^A\langle \tilde{1}\tilde{2}\rangle+\chi_{\tilde{1}}^A\langle \tilde{2}i\rangle+\chi_{\tilde{2}}^A\langle i\tilde{1}\rangle}{\langle i\tilde{1}\rangle\langle \tilde{1}\tilde{2}\rangle},\\
\eta^A_{\widetilde{m+1}}=&{}\frac{\chi_{\tilde{m}}^A\langle \widetilde{m+1}j\rangle+\chi_{\widetilde{m+1}}^A\langle j\tilde{m}\rangle+\chi_{j}^A\langle \tilde{m}\widetilde{m+1}\rangle}{\langle \tilde{m}\widetilde{m+1}\rangle\langle \widetilde{m+1}j\rangle},
\end{align}
\noindent which gives the following super-momentum conserving delta function:

\begin{align}
\delta^4(\chi_{\tilde{1}}^A-\langle\tilde{1}\theta_i^A\rangle)\delta^4(\chi_{\widetilde{m+1}}^A-\langle\widetilde{m+1}\theta_j^A\rangle)&=\left(\frac{\langle ij\rangle}{\langle i\tilde{1}\rangle\langle\widetilde{m+1}j\rangle}\right)^4\delta^8\left(\sum_{r=1}^{m+1}\eta_{\tilde{r}}\lambda_{\tilde{r}}+\sum_{s=j}^{i-1}\eta_s\lambda_s\right)
\end{align}

The factor in front of the delta function cancels part the Jacobian that arises when changing the measure for the $\chi^A$ variables into the measure for the $\eta^A$ variables which is given by:

\begin{align}
\left(\frac{\langle ij\rangle}{\langle i\tilde{1}\rangle\langle\tilde{1}\tilde{2}\rangle\cdots\langle\widetilde{m+1}j\rangle}\right)^4.
\end{align}


\begin{thebibliography}{99}


\bibitem{Bern:1994zx}
  Z.~Bern, L.~J.~Dixon, D.~C.~Dunbar and D.~A.~Kosower,
  Nucl.\ Phys.\ B {\bf 425} (1994) 217
  [hep-ph/9403226].

\bibitem{Bern:1996je}
  Z.~Bern, L.~J.~Dixon and D.~A.~Kosower,
  Ann.\ Rev.\ Nucl.\ Part.\ Sci.\  {\bf 46} (1996) 109
  [hep-ph/9602280].

\bibitem{Britto:2004nc}
  R.~Britto, F.~Cachazo and B.~Feng,
  Nucl.\ Phys.\ B {\bf 725} (2005) 275
  [hep-th/0412103].

\bibitem{Bern:2011qt}
  Z.~Bern and Y.~t.~Huang,
  J.\ Phys.\ A {\bf 44} (2011) 454003
  [arXiv:1103.1869 [hep-th]].

\bibitem{Engelund:2012re}
  O.~T.~Engelund and R.~Roiban,
  JHEP {\bf 1303} (2013) 172
  [arXiv:1209.0227 [hep-th]].

\bibitem{Brandhuber:2010ad}
  A.~Brandhuber, B.~Spence, G.~Travaglini and G.~Yang,
  JHEP {\bf 1101} (2011) 134
  [arXiv:1011.1899 [hep-th]].

\bibitem{Brandhuber:2011tv}
  A.~Brandhuber, O.~Gurdogan, R.~Mooney, G.~Travaglini and G.~Yang,
  JHEP {\bf 1110} (2011) 046
  [arXiv:1107.5067 [hep-th]].

\bibitem{Intriligator:1998ig}
  K.~A.~Intriligator,
  Nucl.\ Phys.\ B {\bf 551} (1999) 575
  [hep-th/9811047].

\bibitem{Laenen:2014jga}
  E.~Laenen, K.~J.~Larsen and R.~Rietkerk,
  arXiv:1410.5681 [hep-th].

\bibitem{Eden:2010zz}
  B.~Eden, G.~P.~Korchemsky and E.~Sokatchev,
  JHEP {\bf 1112} (2011) 002
  [arXiv:1007.3246 [hep-th]].

\bibitem{Eden:2010ce}
  B.~Eden, G.~P.~Korchemsky and E.~Sokatchev,
  Phys.\ Lett.\ B {\bf 709} (2012) 247
  [arXiv:1009.2488 [hep-th]].

\bibitem{Eden:2011yp}
  B.~Eden, P.~Heslop, G.~P.~Korchemsky and E.~Sokatchev,
  Nucl.\ Phys.\ B {\bf 869} (2013) 329
  [arXiv:1103.3714 [hep-th]].

\bibitem{Eden:2011ku}
  B.~Eden, P.~Heslop, G.~P.~Korchemsky and E.~Sokatchev,
  Nucl.\ Phys.\ B {\bf 869} (2013) 378
  [arXiv:1103.4353 [hep-th]].



\bibitem{Alday:2007hr}
  L.~F.~Alday and J.~M.~Maldacena,
  JHEP {\bf 0706} (2007) 064
  [arXiv:0705.0303 [hep-th]].
	
\bibitem{Alday:2007he}
  L.~F.~Alday and J.~Maldacena,
  JHEP {\bf 0711} (2007) 068
  [arXiv:0710.1060 [hep-th]].

\bibitem{Drummond:2007aua}
  J.~M.~Drummond, G.~P.~Korchemsky and E.~Sokatchev,
  Nucl.\ Phys.\ B {\bf 795} (2008) 385
  [arXiv:0707.0243 [hep-th]].

\bibitem{Brandhuber:2007yx}
  A.~Brandhuber, P.~Heslop and G.~Travaglini,
  Nucl.\ Phys.\ B {\bf 794} (2008) 231
  [arXiv:0707.1153 [hep-th]].
	
\bibitem{Mason:2010yk}
  L.~J.~Mason and D.~Skinner,
  JHEP {\bf 1012} (2010) 018
  [arXiv:1009.2225 [hep-th]].
	
	
\bibitem{CaronHuot:2010ek} 
  S.~Caron-Huot,
  JHEP {\bf 1107}, 058 (2011)
  [arXiv:1010.1167 [hep-th]].
	
\bibitem{Belitsky:2011zm}
  A.~V.~Belitsky, G.~P.~Korchemsky and E.~Sokatchev,
  Nucl.\ Phys.\ B {\bf 855} (2012) 333
  [arXiv:1103.3008 [hep-th]].
	
\bibitem{Alday:2010zy}
  L.~F.~Alday, B.~Eden, G.~P.~Korchemsky, J.~Maldacena and E.~Sokatchev,
  JHEP {\bf 1109} (2011) 123
  [arXiv:1007.3243 [hep-th]].
	
\bibitem{Adamo:2011dq}
  T.~Adamo, M.~Bullimore, L.~Mason and D.~Skinner,
  JHEP {\bf 1108} (2011) 076
  [arXiv:1103.4119 [hep-th]].



\bibitem{Alday:2011ga}
  L.~F.~Alday, E.~I.~Buchbinder and A.~A.~Tseytlin,
  JHEP {\bf 1109} (2011) 034
  [arXiv:1107.5702 [hep-th]].

\bibitem{Engelund:2011fg}
  O.~T.~Engelund and R.~Roiban,
  JHEP {\bf 1205} (2012) 158
  [arXiv:1110.0758 [hep-th]].

\bibitem{Adamo:2011cd}
  T.~Adamo,
  JHEP {\bf 1112} (2011) 006
  [arXiv:1110.3925 [hep-th]].

\bibitem{Bork:2011cj}
  L.~V.~Bork, D.~I.~Kazakov and G.~S.~Vartanov,
  JHEP {\bf 1110} (2011) 133
  [arXiv:1107.5551 [hep-th]];
	L.~V.~Bork,
  JHEP {\bf 1301} (2013) 049
  [arXiv:1203.2596 [hep-th]].

\bibitem{Gehrmann:2011xn}
  T.~Gehrmann, J.~M.~Henn and T.~Huber,
  JHEP {\bf 1203} (2012) 101
  [arXiv:1112.4524 [hep-th]].


\bibitem{Brandhuber:2012vm}
  A.~Brandhuber, G.~Travaglini and G.~Yang,
  JHEP {\bf 1205} (2012) 082
  [arXiv:1201.4170 [hep-th]].
	
\bibitem{Johansson:2012zv}
  H.~Johansson, D.~A.~Kosower and K.~J.~Larsen,
  Phys.\ Rev.\ D {\bf 87} (2013) 2,  025030
  [arXiv:1208.1754 [hep-th]].


\bibitem{Young:2013hda}
  A.~Brandhuber, Ö.~Gürdoğan, D.~Korres, R.~Mooney and G.~Travaglini,
  JHEP {\bf 1311} (2013) 022
  [arXiv:1305.2421 [hep-th]];
  D.~Young,
  JHEP {\bf 1306} (2013) 049
  [arXiv:1305.2422 [hep-th]];
  L.~Bianchi and M.~S.~Bianchi,
  Phys.\ Rev.\ D {\bf 89} (2014) 12,  125002
  [arXiv:1311.6464 [hep-th]].

\bibitem{Penante:2014sza}
  B.~Penante, B.~Spence, G.~Travaglini and C.~Wen,
  JHEP {\bf 1404} (2014) 083
  [arXiv:1402.1300 [hep-th]].

\bibitem{Nandan:2014oga}
  D.~Nandan, C.~Sieg, M.~Wilhelm and G.~Yang,
  arXiv:1410.8485 [hep-th];
  M.~Wilhelm,
  arXiv:1410.6309 [hep-th].







\bibitem{Britto:2004ap}
  R.~Britto, F.~Cachazo and B.~Feng,
  Nucl.\ Phys.\ B {\bf 715} (2005) 499
  [hep-th/0412308].
	
\bibitem{Britto:2005fq}
  R.~Britto, F.~Cachazo, B.~Feng and E.~Witten,
  Phys.\ Rev.\ Lett.\  {\bf 94} (2005) 181602
  [hep-th/0501052].


\bibitem{Beisert:2002bb}
  N.~Beisert, C.~Kristjansen, J.~Plefka, G.~W.~Semenoff and M.~Staudacher,
  Nucl.\ Phys.\ B {\bf 650} (2003) 125
  [hep-th/0208178].














\bibitem{Eden:1998hh}
  B.~Eden, P.~S.~Howe, C.~Schubert, E.~Sokatchev and P.~C.~West,
  Nucl.\ Phys.\ B {\bf 557} (1999) 355
  [hep-th/9811172].

\bibitem{GonzalezRey:1998tk}
  F.~Gonzalez-Rey, I.~Y.~Park and K.~Schalm,
  Phys.\ Lett.\ B {\bf 448} (1999) 37
  [hep-th/9811155].

\bibitem{Eden:2000mv}
  B.~Eden, C.~Schubert and E.~Sokatchev,
  Phys.\ Lett.\ B {\bf 482} (2000) 309
  [hep-th/0003096].

\bibitem{Bianchi:2000hn}
  M.~Bianchi, S.~Kovacs, G.~Rossi and Y.~S.~Stanev,
  Nucl.\ Phys.\ B {\bf 584} (2000) 216
  [hep-th/0003203].

\bibitem{Eden:2011we}
  B.~Eden, P.~Heslop, G.~P.~Korchemsky and E.~Sokatchev,
  Nucl.\ Phys.\ B {\bf 862} (2012) 193
  [arXiv:1108.3557 [hep-th]].

\bibitem{Chicherin:2014uca}
  D.~Chicherin, R.~Doobary, B.~Eden, P.~Heslop, G.~P.~Korchemsky, L.~Mason and E.~Sokatchev,
  arXiv:1412.8718 [hep-th].

\bibitem{Binosi:2003yf}
  D.~Binosi and L.~Theussl,
  Comput.\ Phys.\ Commun.\  {\bf 161} (2004) 76
  [hep-ph/0309015].

\end{thebibliography}
\end{document}